\documentclass[reprint,nofootinbib,showpacs,showkeys,prd]{revtex4-1}
\usepackage{graphicx}
\usepackage{epsfig}
\usepackage{amsmath}
\usepackage{amssymb}
\usepackage{calc}
\begin{document}
\title{Berry curvature, horocycles and scattering states in $AdS_3/CFT_2$}
\author{P\'eter L\'evay}
\affiliation{MTA-BME Quantum Dynamics and Correlations Research Group, Department of Theoretical Physics, 
Budapest University of  Technology and Economics, 1521 Budapest, Hungary}

\date{\today}

\begin{abstract}
By studying the space of geodesics in $ADS_3/CFT_2$ and quantizing the geodesic motion, we relate scattering data to boundary entanglement of the CFT vacuum.
The basic idea is to use a family of plane waves parametrized by coordinates of the space of geodesics i.e. kinematic space.
This idea enables a simple calculation of the Berry curvature living on kinematic space.
As a result we recover the Crofton form with a coefficient depending on the scattering energy.
In arriving at these results the space of horocycles is used. 
We show that this new space used in concert with kinematic space incorporates naturally the gauge degrees of freedom responsible for an analogue of Berry's Phase.
Horocycles also give a new geometric look to the strong subadditivity relation in terms of lambda lengths giving rise to shear coordinates of geodesic quadrangles.
A generalization for geodesic polygons then reveals an interesting connection with $A_{n}$ cluster algebras. Here the cluster variables are the lambda lengths related to the regularized entropies of the boundary via the Ryu-Takayanagi relation.
An elaboration of this idea indicates that
cluster algebras might provide
a natural algebraic means for encoding the gauge invariant entanglement patterns of certain boundary entangled states in the geometry of bulk geodesics.
Finally using the language of integral geometry we show how certain propagators connected to the bulk, boundary and kinematic spaces
are related to data of elementary scattering problems.
We also present some hints how these ideas might be generalized for more general holographic scenarios.
\end{abstract}


\maketitle

\section{Introduction}

Due to the progress in the field of AdS/CFT it has become clear by now that entanglement patterns of certain quantum states associated to one type of space (boundary) are encoded into geometric structures of an other one (bulk)\cite{RT,RT2,HRT,Raam1,Raam2,Raam3}.
Much recently a new type of space, the space of geodesics called kinematic space, has been invoked in the hope to act as an interpreter between the original spaces via the universal language of integral geometry\cite{Czech1,Czech1b}.
In the simplest case of $AdS_3$ this kinematic space is of the product form of two copies of two dimensional de Sitter spaces\cite{Czech1b}. A further specification to the kinematic space of geodesics on a space-like slice reveals a connection between an asymptotically anti-de Sitter bulk and a single copy of a two dimensional de Sitter space\cite{Czech1}. 
Now combining this with the original $AdS_3/CFT_2$ correspondence this picture gives rise to the idea of an emergent de Sitter from conformal field theory.
This idea has also shown up in yet another context from the first law of entanglement entropy for perturbations around the vacuum state of CFTs\cite{Myers}.
 
The key element in this new bulk/kinematic space correspondence 
is a smooth manifold (bulk) together with a family of its submanifolds (geodesics) parametrized by the points of another manifold (kinematic space).
Hence in this setup
the points of kinematic space correspond to the lines (geodesics) of the bulk.
However, the points of the bulk also parametrize a family of submanifolds (point curves) of kinematic space\cite{Czech1}.
This duality between these spaces can be grasped in a mathematically precise manner by the notion of a double fibration\cite{Chern,Helga3}.

This story has already been well-known for physicists familiar with the basic correspondence of twistor theory\cite{Penrose}. There the points of a space (four dimensional compactified and complexified Minkowski space-time) are related to the lines of another one (the three dimensional complex projective twistor space). Then integral geometry helps to relate data on both sides of the correspondence.
However, apart from a single double fibration in twistor theory what is also considered is a {\it set of double fibrations} used in concert by adjoining further spaces forming further double fibrations\cite{WardWells}.
Following this idea in this paper we would like to drew the readers attention to the utility of the space of horocycles in a holographic context. Taking together with our bulk this new space is forming yet another double fibration. 
The aim of this paper is to illustrate the physical meaning of such a mathematical structure in an elementary manner via the simplest example one can have, namely in the $AdS_3/CFT_2$ setup. Our hope is that these simple observations might whet the reader's appetite to explore this idea further for more general holographic scenarios.

The physical basis which connects the space of geodesics to the space of horocycles is scattering theory.
As is well-known, scattering theory is naturally associated with states described by a convenient set of asymptotic (boundary) data.
On the other hand, classically the domain of interaction (bulk) serves as a region to be probed by test particles following geodesics whose characteristics are determined precisely by such asymptotic data.
In this scattering language horocycles act like monitoring stations where the test particles following the geodesics are registered after (or before) being scattered i.e. probing the deeper regions of the bulk. The arbitrariness in the choice of a horocycle corresponds to an ambiguity for choosing a gauge. 
In the simplest case of pure $AdS_3$ the scattering problem is associated with a mathematical description featuring some scattering potential accounting for such a domain of interaction. However, since this type of scattering is purely geometric in origin 
for more general examples of $2+1$ dimensional gravity no description based on potentials is expected to be available.

The advantage of the viewpoint provided by scattering theory is that
it is also possible to consider the quantization of the classical geodesic motion, resulting in some quantum mechanical scattering problem encapsulating new messages about holography.
In the special case of taking the static slice in  $AdS_3/CFT_2$ resulting in the Poincar\'e disc $\mathbb D$ factorized by some discrete subgroup $\Gamma$ of its group of isometries $G=SU(1,1)$, the resulting quantum scattering problems are the ones already familiar from the literature. Indeed, this is the topic of scattering theory of automorphic functions\cite{Fad,LP} connected to the literature on quantum chaos\cite{Gutz,BV,Comtet,Pnueli}. 
On the other hand as is well-known a large variety of spacetimes, including the BTZ black hole\cite{BTZ} can be obtained by factorizing $AdS_3$ by means of similar discrete groups of isometries\cite{Ingemar1,Brill}.
In interesting special cases\cite{Skenderis} one can obtain this factorization by extending the action of the isometries of $\mathbb D$ to isometries on $AdS_3$. 
Moreover, scattering situations familiar from the chaotic scattering literature featuring cusps, can in principle be embedded into these scenarios via extremal black holes\cite{Ingemar1,Brill}.
This scattering language has also been used for expressing various thermodynamic quantities of the BTZ black hole in terms of the Selberg zeta function\cite{Perry,Aros,Aros2}.

In this paper we would like to draw the readers attention to some new perspectives this scattering based viewpoint can offer for issues of holography, kinematic space and integral geometry. Though what we are having in mind for future work is to explore the interesting possibilities which are lying in scattering situations where the bulk spacetime is of the form $AdS_3/\Gamma$, here we will be content with some simple scattering based observations on the geometry of pure $AdS_3$ dual to the vacuum of a $CFT_2$. 

Namely we will consider the static slice of $AdS_3$ which is the Poincar\'e disc $\mathbb D$, and introduce two dual spaces. One of the them is the space of geodesics (kinematic space) $\mathbb K$ and the other is the space of horocycles (the space of regularizators for geodesics)  $\mathbb G$ of $\mathbb D$.
The variables of $\mathbb D$ and $\mathbb K$ will be regarded as fast and slow ones reminiscent of the usual splitting of dynamical variables in the adiabatic scenario familiar from the Berry's Phase literature\cite{Berry,SW}. 
On the other hand  $\mathbb G$ can also be regarded as the space of gauge choices, i.e. a space parametrizing a local $GL(1,\mathbb C)$ degree of freedom, an analogue of Berry's Phase for scattering states.

The organization of this paper is as follows.
In Section II. we hint at the possibility of relating scattering plane waves to the data of kinematic space, and introduce horocycles in this context.
In Section III. we elaborate on this idea by introducing a geodesic operator giving rise to a family of scattering states parametrized by the points of kinematic space. An elementary argument of Section IV. shows that the Berry curvature for this family should be proportional to the Crofton form on $\mathbb K$ with a coefficient depending on the scattering energy. 
We will show that the gauge degree of freedom associated with the appearance of a $GL(1,\mathbb C)$ version of Berry's Phase
is connected to the space of horocycles $\mathbb G$. A convenient way of looking at this space is as a bulk manifestation of the "space of cutoffs" in the boundary $\partial\mathbb D$ in a geometric manner.
These results can be considered as an elaboration on the ideas that showed up first in Ref.\cite{Czech2}.

In Section V. we write the scattering wave function as a superposition of plane waves emitted from all boundary points.
We have two density distributions on the boundary, one of them corresponding to the starting and end points of the geodesics respectively. Then our parametrized family of plane waves can be regarded as boundary to bulk propagators giving rise to the scattering wave function of the bulk.
The two distributions are shown to be related by a scattering operator acting as an intertwiner an idea familiar from the literature on Algebraic Scattering Theory (AST)\cite{Kerimov,Kerimov2,Iachello}.
The phase of the kernel of the scattering operator can be connected to a quantity analogous to the Wigner delay known from scattering theory. 
In the special case of pure $AdS_3$ we observe that the Wigner delay is proportional to the entanglement entropy with a cutoff depending on the scattering energy. 

In Section VI. using horocycles and their associated lambda lengths\cite{Penner,Pennerbook} we give a new geometric meaning to the well-known strong subadditivity relation for entanglement entropies.
In particular we relate two aggregate measures on the boundary detecting how far the infrared degrees of freedom are away from satisfying the strong subadditivity relation to the shear coordinates of geodesic quadrangles in the bulk. This boundary measure, the conditional mutual information is simply related to the geodesic distance between the geodesics forming the opposite sides of the quadrangle, ones having time-like separation as points in kinematic space.
In the dual picture provided by kinematic space the conditional mutual information is also related to the proper time along a timelike geodesic connecting these two points.

In Section VII. we draw the readers attention in this context to a connection with the theory of Teichm\"uller spaces of marked Riemann surfaces. In particular we show that the two different boundary measures of strong subadditivity are related to the two different triangulations of a geodesic quadrangle. In this picture the shear coordinates, satisfying a reciprocial relation, are just possible local coordinates for the space of deformations of quadrangles, i.e. their Teichm\"uller space.
Generalizing these observations for geodesic $n$-gons with $n\geq 4$ an interesting connection with $A_{n-3}$ cluster algebras\cite{Williams} emerges.  Here the cluster variables are just the lambda lengths of Penner\cite{Penner} directly related to the regularized entropies of the boundary via the Ryu-Takayanagi relation.
We also emphasize that the basic role the gauge invariant conditional mutual informations play in these elaborations
dates back to the unifying role of the Ptolemy identity (\ref{ptolemy}).
This identity is the basis of recursion relations underlying transformation formulas for shear coordinates of geodesic pentagons.
Indeed in the general case of geodesic $n$-gons these recursion relations are precisely of the form of Zamolodchikov's $Y$-systems of $A_{n-3}$ type\cite{Zamo}.
Moreover, since boundary intervals with their associated geodesics of the bulk are organized according to the causal 
structure of their corresponding points in kinematic space\cite{Czech1}, in this manner the underlying quivers of cluster algebras are connected to structures of causality.

Finally in Section VIII. using the language of integral geometry we show how certain propagators of the literature connected to our spaces $\mathbb D$, $\mathbb K$ ,$\partial\mathbb D$
are related to elementary scattering problems.
Some comments, speculations on further topics of exploration and the conclusions are left for Section IX. 
For the convenience of the reader Appendix A. contains the basic definitions on horocycles and lambda lengths, and Appendix B. contains the definitions of the double fibrations, structures used implicitely in our elaborations.

\section{Plane waves and kinematic space}

We introduce coordinates for the spacelike slice of $AdS_3$ as
\begin{subequations}
\begin{align}
X\equiv X_1&=\sinh\varrho\cos\varphi,\\
Y\equiv X_2&=\sinh\varrho\sin\varphi,\\
U\equiv X_3&=\cosh\varrho,
\label{XYZ}
\end{align}
\end{subequations}
with $X^2+Y^2-U^2=-R^2,\quad R=1$.
This is just the upper sheet ${\mathbb H}$ of the double sheeted hyperboloid that can also be written as the coset space
$\mathbb H\simeq SO(2,1)/SO(2)$.
After stereographic projection of ${\mathbb H}$ to the Poincar\'e disc ${\mathbb D}$ we obtain the coordinates
\begin{equation}
z=\tanh({\varrho}/2)e^{i{\varphi}}=\frac{X+iY}{1+U}=x+iy\in{\mathbb D}.
\label{ze}
\end{equation}
An alternative set of coordinates can be obtained by transforming to the upper half plane $\mathbb U$ by a Cayley transformation
\begin{equation}
\tau=i\frac{1+z}{1-z}=\frac{i-Y}{U-X}=\xi+i\eta\in{\mathbb U},\qquad \eta>0.
\label{tau}
\end{equation}
On ${\mathbb H}$ we have the metric $ds^2=d{\varrho}^2+\sinh^2\varrho d\varphi^2$ with its geodesics given by the formula
\begin{equation}
\tanh\varrho\cos(\varphi-\theta)=\cos\alpha.
\label{geo}
\end{equation}
Here the extra parameters $\theta\in[0,2\pi]$ and $\alpha\in [0,\pi]$ are labelling the geodesics.
Depicted on the disc $\mathbb D$ the coordinate $\theta$ is the center and $\alpha$ is half the opening angle of the geodesic see Figure 1.
\begin{figure}
\centerline{\includegraphics[width=6truecm,clip=]{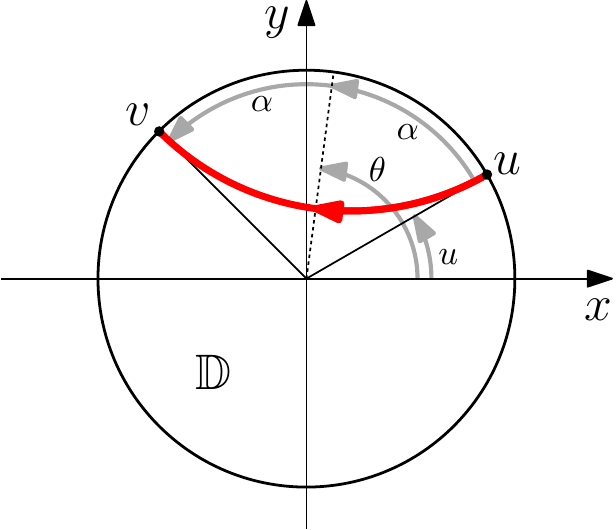}}
\caption{The parametrization of a geodesic. The geodesic (red curve) is parametrized by the pair
$(\theta,\alpha)$ where  $\theta\in[0,2\pi]$ and $\alpha\in [0,\pi]$. The coordinate $\theta$ is the center and $\alpha$ is half the opening angle of the geodesic. Alternatively one can use the pair $(u,v)$ defined by Eq.(\ref{uv}).}
\end{figure}
Pairs of geodesics differing in orientation are related by $\theta\leftrightarrow \theta+\pi$, $\alpha\leftrightarrow \pi-\alpha$.
Hence our space of geodesics is labelled by the coordinates $(\alpha, \theta)$. It is called the kinematic space\cite{Czech1}.
Topologically the kinematic space is the single sheeted hyperboloid $\mathbb K\simeq SO(2,1)/SO(1,1)$ which is the deSitter space $dS_2$. 

We are interested in scattering states arising from the quantization of the geodesic motion on $\mathbb D$.
More precisely what we would like to get is a family of scattering states parametrized by the coordinates of the points in kinematic space $\mathbb K$. 
In the language of the literature on Berry's Phase\cite{Berry,SW} we would like to regard the coordinates $(\varrho,\varphi)$ labelling points in $\mathbb H$ as fast variables (to be quantized), and the ones $(\alpha, \theta)$ labelling points in $\mathbb K$ as slow ones (regarded as parameters).

As is well-known\cite{BV} quantization of the geodesic motion of a particle with mass $2\mu=1$ and energy $E$ on $\mathbb H$ (i.e. on the "pseudosphere" of unit radius $R=1$) in the scattering domain is effected by considering solutions of the equation
\begin{equation}\label{quanteq}
\left(\triangle+\frac{1}{4}+k^2\right)\psi=0
\end{equation}
with
\begin{equation}
\triangle=\frac{1}{\sinh\varrho}\frac{\partial}{\partial\varrho}\left(\sinh\varrho\frac{\partial}{\partial\varrho}\right)+\frac{1}{\sinh^2\varrho}\frac{\partial^2}{\partial\varphi^2}.
\end{equation}
Here $\triangle$ is the Laplacian on $\mathbb H$ which can be written as the quadratic Casimir of the group $SO(2,1)$. Scattering states are given by the irreps of $SO(2,1)$ belonging to the continuous principal series of irreps labelled by the komplex number $j=-1/2+ik$, with $k\in{\mathbb R}^+$ and $1/4+k^2=E$ ($\hbar=1$) being the scattering energy.  
In this notation the Hamiltonian is $H=-\triangle$ and its eigenvalue is of the form $-j(j+1)=1/4+k^2$.
Notice that after putting back all the constants $(R,\mu,\hbar)$ we get
\begin{equation}
E=\frac{\hbar^2}{2\mu R^2}(1/4+k^2)
\end{equation}
hence the semiclassical limit $\hbar\to 0$ corresponds to the one of $1/4+k^2 \to \infty$. Hence for large $k$ one can regard
$1/k$ as a semiclassical parameter\cite{BV}. 

In the upper half plane realization $\triangle$ can be written as $\triangle=\eta^2(\partial^2_{\xi}+{\partial}^2_{\eta})$.
Then one can check that the functions on $\mathbb U$ of the form
\begin{equation}
\psi_{0k}^{\pm}(\eta)\equiv \eta^{\frac{1}{2}\pm ik}=\sqrt{\eta}e^{\pm ik\log\eta}
\label{base}
\end{equation}
\noindent 
satisfy Eq.(\ref{quanteq}). 
One can also verify\cite{BV,Gutz} that these solutions $\psi_{0k}^{\pm}(\eta)$ correspond to the semiclassical (WKB) wave functions interpreted as plane waves going to (respectively coming from) the boundary point $\eta=\infty$.
In the disc model $\mathbb D$ this sink (or source) of plane waves is the point with coordinates: $(x,y)=(1,0)$.
Hence for example the plane wave 
$\psi_{0k}^{-}(\eta)$
sourced at  $(x,y)=(1,0)$ is associated with the geodesic starting at  $\varphi_-\equiv u =0$ going through the origin $z_0=0$ and then arriving at $\varphi_+\equiv v =\pi$. This geodesic is a diametrical one. Its parameters are characterized by the kinematic space coordinates: $(\alpha,\theta)=(\pi/2,\pi/2)$. 
One can alternatively use the coordinates $(u,v)=(0,\pi)$ of Figure 1. related to the pair $(\alpha,\theta)$
as
\begin{equation}
u=\theta-\alpha,\qquad v=\theta +\alpha.
\label{uv}
\end{equation}

Now since, $\eta=(U-X)^{-1}=(\cosh\varrho-\sinh\varrho\cos\varphi)^{-1}$ one can write our outgoing and incoming plane waves as
\begin{equation}
\psi_{0k}^{\pm}(\varrho,\varphi;0,\pi)=(U-X)^{-1/2\mp ik}.
\label{base1}
\end{equation}
The notation on the left hand side emphasizes the role of the variables which they will play in our forthcoming considerations. Namely, the pair $(\varrho,\varphi)\in {\mathbb H}$ corresponds to the fast, and $(u,v)=(0,\pi)\in{\mathbb K}$ to the slow variables.

Let us also refer to an alternative parametrization, which also elucidates the geometric meaning of our plane waves. 
In this new parametrization we use the pair $(u,z_0)$ where $z_0$ is a {\it special point} of ${\mathbb D}$. Hence our geodesic is starting from the boundary point $u$ and going through the specially chosen bulk one $z_0=\tanh(\varrho_0/2)e^{i\varphi_0}$. Clearly $(u,z_0)$ determines the endpoint coordinate $v$.
Indeed, a calculation shows that we have the formula
\begin{equation}
\tan\left(\frac{v-u}{2}\right)=\frac{\tanh\varrho_0\cos(u-\varphi_0)-1}{\tanh\varrho_0\sin(u-\varphi_0)}.
\label{rel}
\end{equation}
Notice that form Eq.(\ref{rel}) the formula obtained in Eq.(4.6) of Ref.\cite{Czech1} follows, namely
\begin{equation}
\cos\alpha=\tanh\varrho_0\cos(\varphi_0-\theta)
\end{equation}
where the parameters $(\varrho_0,\varphi_0)$ of $z_0$ determine the magnitude and direction of the boost needed to move the geodesic with parameters $(\alpha,\theta)=(\pi/2,\pi/2)$ to a one with an arbitrary 
$(\alpha,\theta)$.

Apart from having the meaning as a point encapsulating boost parameters, $z_0$ also has a nice physical interpretation related to scattering theory.
Indeed, let us form the {\it horocycle}\footnote{For the algebraic definition of horocycles and some of their properties see Appendix A.} at the boundary point $\omega\equiv e^{iu}$ which goes through $z_0$.
A horocycle $h$ at $\omega \in \partial{\mathbb D}$ is a Euclidean circle in ${\mathbb D}$ tangent at $\omega$ to $\partial{\mathbb D}$. The point $\omega$ will be called the base point of the horocycle.
Then in the case when $(u,v)=(0,\pi)$ and $z_0=0$ 
we have 
\begin{equation}
\psi_{0k}^{\pm}(z;0,\pi)\equiv
e^{(1/2\pm ik)d(\omega;z,0)}
\label{diaplane}
\end{equation}
where
$z=\tanh(\varrho/2)e^{i\varphi}$ and
$d(\omega;z,z_0)$ is the hyperbolic length of that part of the geodesic departing from $\omega=1$ which is between $z_0=0$ and the horocycle which is based at the point of arrival $\omega=-1$ and goes through $z$, see Figure 2. Notice that the horocycles are acting as natural regularizators for the geodesics having infinite hyperbolic length.
According to Gutzwiller\cite{Gutz} the horocycles can also be regarded as some sort of monitoring stations where the test particles following the geodesics are registered after (and before) they are being scattered, i.e. probing the deeper regions of the bulk.
\begin{figure}
\centerline{\includegraphics[width=\columnwidth]{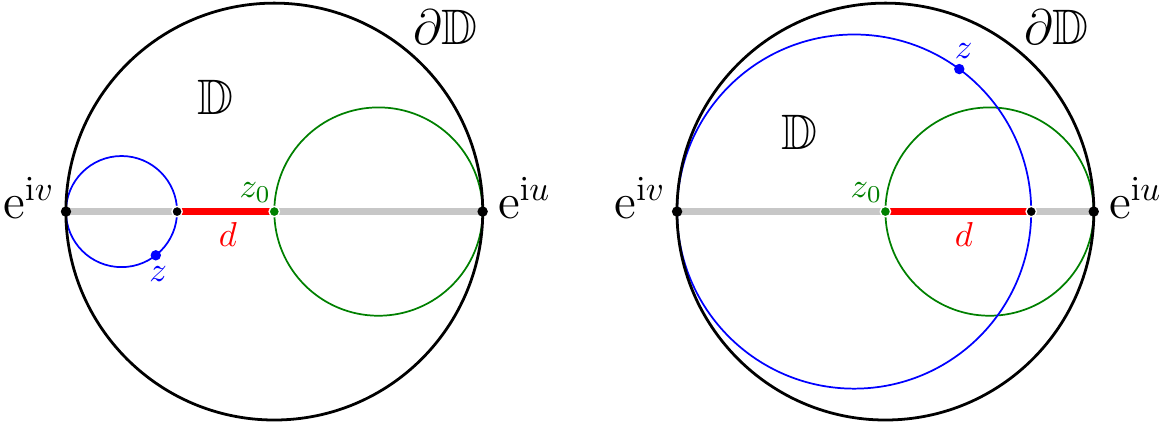}}
\caption{The meaning of our (\ref{diaplane}) plane wave associated to a diametrical geodesic departing from $\omega=e^{iu}=1$ and going through the point $z_0=0$.
The green and blue horocycles going through $z_0$ and $z$ are either non intersecting (left) or intersecting (right). In the respective cases the hyperbolic length $d(\omega;z,z_0)$ of the red geodesic segment is positive (left) or negative (right).}
\end{figure}

\section{A geodesic operator of the Berry form}

In this section our aim is to construct an operator associated to an arbitrary geodesic of the spacial slice of global $AdS_3$ such that its eigenfunctions describe the corresponding plane waves.
In order to do this let us introduce
 the generators $J_1,J_2,J_3$ of $SO(2,1)$. These differential operators are the usual Killing vectors of ${\mathbb H}$ expressed in terms of the fast variables $(\varrho,\varphi)$
\begin{equation*}
J_1\pm iJ_2=e^{\pm i\varphi}\left(\mp\partial_{\varrho}-i\coth\varrho\partial_{\varphi}\right),\qquad J_3=-i\partial_{\varphi}
\end{equation*}
satisfying the commutation relations
\begin{equation}
[J_1,J_2]=-iJ_3, \quad [J_2,J_3]=iJ_1,\quad [J_3,J_1]=iJ_2.
\label{so21kommut}
\end{equation}
Since the kinematic space $\mathbb K$ described by the coordinates $(\alpha,\theta)$ is a de Sitter space $dS_2$ one can also use the coordinates
\begin{subequations}\label{haha}
\begin{align}
B_1&=\frac{\cos\theta}{\sin\alpha}=\cosh\gamma\cos\theta\\,\qquad B_2&=\frac{\sin\theta}{\sin\alpha}=\cosh\gamma\sin\theta\\, \qquad B_3&=\cot\alpha=-\sinh\gamma
\end{align}
\end{subequations}
where the relation $B_1^2+B_2^2-B_3^2=1$ holds, and we introduced the coordinate transformation 
$\cosh\gamma=1/{\sin\alpha}$.

Now our geodesic operator is defined as
\begin{equation}
\mathcal{H}({\bf B})\equiv {\bf J}\cdot{\bf B}=J_1B_1+J_2B_2-J_3B_3.
\label{hamiltoni}
\end{equation}
This operator is just an $SO(2,1)$ analogue of a Hamiltonian for a spin (fast variable) coupled to a magnetic field (slow variable) showing up in studies of Berry's Phase\cite{Berry,SW}. 
Clearly we have
\begin{equation}\label{ittis}
[\mathcal{H}({\bf B}),{\bf J}+{\bf K}]=0
\end{equation}
where ${\bf K}=(K_1,K_2,K_3)$ are the Killing vectors of ${\mathbb K}$
with explicit form
\begin{equation*}
K_1\pm iK_2=e^{\pm i\theta}\left(\mp\partial_{\gamma}-i\tanh\gamma\partial_{\theta}\right),\quad K_3=-i\partial_{\theta}.
\end{equation*}
Notice that  Eq.(\ref{geo}) for our geodesics can alternatively be described by the constraint
${\bf X}\cdot{\bf B}=0$, i.e. the vectors ${\bf X}$ and ${\bf B}$ are Minkowski orthogonal, with the former is a time-like and the latter is a space-like unit vector. 
As has been demonstrated in Ref.\cite{BV} the vector ${\bf B}$ can be regarded as the vector of conserved quantities for the geodesic motion on the pseudosphere
${\mathbb H}$.

Now we have $J_2=i(U{\partial}_X+X{\partial}_U)$ which is a boost. This operator shows up in $\mathcal{H}({\bf B})$ for ${\bf B}={\bf B}_0\equiv (0,1,0)$. Let us denote this operator as 
\begin{equation}
\mathcal{H}({\bf B}_0)\equiv \mathcal{H}_0=J_2.
\end{equation}
Clearly we have
\begin{equation}
{\mathcal H}_0\psi_{0k}^{\pm}=\kappa_{\pm}\psi_{0k}^{\pm},\qquad \kappa_{\pm}=\frac{i}{2}\mp k
\label{hanull}
\end{equation}
where $\psi_{0k}^{\pm}$ is given by Eq.(\ref{base1}).
Hence these incoming and outgoing plane-waves are eigenfunctions of ${\mathcal H}_0$.

It is straightforward to check that we have
\begin{equation}
\mathcal{H}({\bf B})=e^{i(\pi/2-\theta) J_3}e^{i\gamma J_1}\mathcal{H}_0 e^{-i\gamma J_1}e^{-i(\pi/2-\theta) J_3}.
\label{coadjoint}
\end{equation}
As a result of this we can associate plane waves to each of our geodesics.
Indeed, as an example
let us chose $\theta = u+\pi/2$, i.e. $\alpha=\pi/2$. As $u$ varies the corresponding geodesics (which are just diameters of ${\mathbb D}$ centered at the point $z_0 =0$) cover all of ${\mathbb D}$. These geodesics correspond to the canonical point curve in ${\mathbb K}$ of Ref.\cite{Czech1}. In this case $\gamma=0$ hence the corresponding family of plane waves associated to this family of geodesics is
\begin{equation}\label{helga}
\begin{split}
\psi_{k}^{\pm}(\varrho,\varphi;u,u+\pi)
& =e^{-iuJ_3} 
\psi_{0k}^{\pm}(\varrho,\varphi;0,\pi)\\
& =(\cosh\varrho-\sinh\varrho\cos(\varphi-u))^{-\frac{1}{2}\mp ik}.
\end{split}
\end{equation}
These plane waves are eigenfunctions of $\mathcal{H}(\bf B)$ with $\theta = u+\pi/2$ and $\alpha=\pi/2$ with the eigenvalue 
$\kappa_{\pm}$.
The most general plane wave is of the form
\begin{equation}
\psi_{k}^{\pm}(\varrho,\varphi;u,v)=e^{-iuJ_3}e^{i\gamma J_1}\psi_{0k}^{\pm}(\varrho,\varphi;0,\pi)
\label{generalplane}
\end{equation}
where
${\sin\left((v-u)/2\right)}=1/\cosh\gamma$.
Yet another way of writing these incoming and outgoing plane waves is
\begin{equation}\label{wavedistance}
\begin{split}
\psi^{\pm}_k(\varrho,\varphi;u,v)& =
e^{(1/2\pm ik)d(\omega;z,z_0)}\\& =\lambda^{1\pm 2ik}(h_+(z),h_-(z_0))
\end{split}
\end{equation}
where
\begin{equation}
z=\tanh\frac{\varrho}{2}e^{i\varphi},\quad
z_0=\tanh\frac{\varrho_0}{2}e^{i\varphi_0},\quad \omega=e^{iu}
\nonumber
\end{equation}
where
$\tanh\gamma=\tanh\varrho_0\cos(\varphi_0-\theta)$.
For the definition of the {\it lambda length} $\lambda(h_+,h_-)$ see Eqs.(\ref{lambdalength})-(\ref{lambdalength2}) of Appendix A.
The meaning of the parameters showing up in these expressions is also explained in Figure 3.

\begin{figure}
\centerline{\includegraphics[width=\columnwidth]{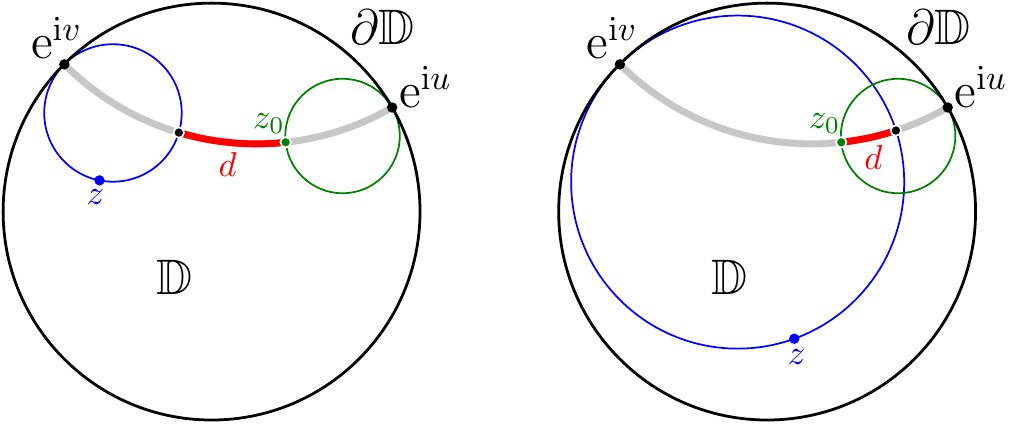}}
\caption{The meaning of our (\ref{wavedistance}) plane waves for an arbitrary spacelike geodesic going through the point $z_0$.
The green and blue horocycles going through $z_0$ and $z$ respectively are either non intersecting (left) or intersecting (right). In the respective cases the distance function $d(\omega;z,z_0)$ associated to the corresponding red geodesic segments is positive (left) or negative (right).}
\end{figure}

\section{Berry curvature}

In this section we calculate the Berry curvature and give an explicit description of the $SO(1,1)$ gauge degree of freedom
manifesting in a $GL(1,\mathbb C)$ version of the usual Berry's Phase.
Our treatise can be regarded as an elaboration on the ideas of Ref.\cite{Czech2}.

First of all notice that the definition of the notion of Berry's curvature rests on the quantum adiabatic theorem for Hamiltonians depending on a set of slowly changing parameters. Slow means that the kinetic energy associated with the classical motion of the parameters can not to induce transitions between the eigenstates belonging to different eigensubspaces. 
This is the case when there is a gap in the spectrum of the Hamiltonian and this gap is much larger than the energy scale associated to slow motion. 
In our case the analogue of the parametrized families of Hamiltonians are the (\ref{hamiltoni}) geodesic operators with the set of parameters belonging to kinematic space $\mathbb K$. 
However, now as eigenstates we have the scattering states belonging to the continuous part of the spectrum so it is not obvious how to generalize these notions in a mathematically precise way.
Such a generalization should be formulated without the use of a gap condition \cite{Avron} and for the continuous spectrum \cite{Maamache} issues already discussed in the literature.
Here we are not addressing the issue of arriving at a precise formalism along this line. We will merely use a purely formal definition of Berry's connection which is sufficient for our considerations.

Let us now recall that it is known that the orthogonality relation for the incoming and outgoing waves of Eq.(\ref{helga}) is\cite{Helgason}
\begin{equation*}
\langle\psi_k^{\pm}\vert\psi_{k^{\prime}}^{\pm}\rangle =(k\tanh\pi k)\delta(k-k^{\prime})\delta(u-u^{\prime})
\end{equation*}
moreover, either the incoming or the outgoing ones form a complete set.
Normalizing by the $k$-dependent factors ($k\in{\mathbb R}^+$) one arrives at a delta-normalized plane wave $\tilde{\psi}$ satisfying
$\langle\tilde{\psi}_{k}^{\pm}\vert\tilde{\psi}_{k^{\prime}}^{\pm}\rangle =\delta(k-k^{\prime})\delta(u-u^{\prime})$.

A formal definition of the Berry conection is
\begin{equation*}
\mathcal{A}^{(\pm)}_{kk^{\prime}}=-\Im\langle \tilde{\psi}_{k}^{\pm}\vert d\vert\tilde{\psi}_{k^{\prime}}^{\pm}\rangle=-\Im\langle\tilde{\psi}_{0k}^{\pm}\vert U^{-1}dU\vert\tilde{\psi}_{0k^{\prime}}^{\pm}\rangle
\end{equation*}
where $\Im$ refers to taking the imaginary part and we have used Eq.(\ref{generalplane}) with $U=e^{i(\pi/2-\theta)J_3}e^{i\gamma J_1}$. 
The exterior derivative acts on the slow variables and $\langle\cdot\vert\cdot\rangle$ refers to integration with respect to the fast variables.
The quantity $ U^{-1}dU$ is the pull-back of the Maurer-Cartan form for the group manifold $SO(2,1)$ to ${\mathbb K}$. It is an $so(2,1)$ Lie-algebra valued one form hence it can be written as a linear combination of the generators $J_1,J_2$ and $J_3$ regarded as basis vectors. The expansion coefficients are one-forms living on ${\mathbb K}$. 
They can be calculated using any faithful matrix representation for the $so(2,1)$ algebra.
For example the one $(J_1,J_2,J_3)=(-\frac{i}{2}\sigma_2,\frac{i}{2}\sigma_1, \frac{1}{2}\sigma_3)$ will do.
Then a calculation shows that $\langle\tilde{\psi}_{0k}^{\pm}\vert J_1\vert\tilde{\psi}_{0k}^{\pm}\rangle=\langle\tilde{\psi}_{0k}^{\pm}\vert J_3\vert\tilde{\psi}_{0k}^{\pm}\rangle =0$, hence only the one-form valued coefficient of $J_2$ is needed.
This yields
\begin{equation*}
\mathcal{A}^{(\pm)}_{kk^{\prime}}=\Im\left(i\langle\tilde{\psi}_{0k}^{\pm}\vert J_2\vert\tilde{\psi}_{0k^{\prime}}^{\pm}\rangle\right) \sinh\gamma d\theta.
\end{equation*}
Now using Eqs.(\ref{haha}) and (\ref{hanull}) one obtains
\begin{equation}\label{Bconn}
\mathcal{A}^{(\pm)}_{kk^{\prime}}=\pm k{\cot\alpha}d\theta\delta(k-k^{\prime}).
\end{equation}
The Berry curvature is
\begin{equation}\label{curvature}
{\mathcal F}^{(\pm)}_{kk^{\prime}}=d\mathcal{A}^{(\pm)}_{kk^{\prime}}=\pm k\frac{d\theta\wedge d\alpha}{\sin^2\alpha}\delta(k-k^{\prime}).
\end{equation}

Now we recall that in the integral geometric approach of Ref.\cite{Czech1} the correct choice for the Crofton form is
\begin{equation}
\omega=\frac{\partial^2S(u,v)}{\partial u\partial v}du\wedge dv
\end{equation}
where
\begin{equation}
S(u,v)=\frac{c}{3}\log\left(e^{\Lambda}\sin\left(\frac{v-u}{2}\right)\right)
\label{Suv}
\end{equation}
with $e^{\Lambda}$ is the cutoff factor.
After using Eq.(\ref{uv}) we see that the Berry curvature is related to the Crofton-form times a factor depending on the scattering energy.
The sign ambiguity showing up in our Berry curvature calculation amounts to the incoming and outgoing representation of plane waves. They correspond to the two different orientations of the underlying geodesics related by the transformation $(\theta,\alpha)\leftrightarrow (\theta+\pi,\pi-\alpha)$. 
In Section V. we show that the incoming and outgoing densities of Eq.(\ref{Wig}) are related by an intertwining kernel whose phase is related to $S(u,v)$ with a $k$-dependent cutoff.

Berry's connection, showing up in our calculation, is also described by the natural Riemannian connection on the symmetric space 
$\mathbb K$.
In fact, it is known\cite{Giler} that for parameter spaces of the homogeneous form $G/H$ the necessary and sufficient condition for the Berry connection to be related to this Riemannian connection is that $G/H$ is a symmetric space and that the matrix elements of the generators not belonging to the Lie algebra of $H$ are vanishing.
In our case both of these conditions hold.

Another consequence of this is that the symmetric part of the quantum geometric tensor\cite{Provost,SW}, which defines a metric
on $\mathbb K$ up to some $k$-dependent factors, should be proportional to this natural Riemannian metric. 
As is well-known when also quantizing the parameters within a Born-Oppenheimer like treatment, the trace of this metric
gives rise to an electric type of gauge force\cite{BLim}. 
In our $AdS_3/CFT_2$ setting the line element $ds^2_{BO}$ of this metric up to a crucial $k$ dependent factor  should be just the one giving rise to the line element of Ref.\cite{Czech1}
\begin{equation}
ds^2=\frac{\partial S(u,v)}{\partial u\partial v}dudv.
\label{PVmetr}
\end{equation}
It would be interesting to calculate this $k$ dependent factor and clarify the meaning of $ds^2_{BO}$ in a holographic context. 
Note that the sign of this factor is crucial, since according to Eq.(\ref{ssshear3}) $ds^2$ is connected to the strong subadditivity of boundary entanglement entropies\cite{Czech1}.

Now we discuss the gauge degree of freedom familiar from the literature on Berry's Phase.
Let us recall our most general (\ref{wavedistance}) eigenstate of the (\ref{hamiltoni}) geodesic operator. 
We are going to show that the gauge degree of freedom can be traced back to our freedom in choosing the green horocycle of Figure 3.
Fixing the point of tangency of this horocycle at $e^{iu}$ this freedom of choice boils down to our choice of base point $z_0$ located along our fixed geodesic. Note that fixing an oriented geodesic amounts to fixing our slow variables $(u,v)$.
Since horocycles at $\omega=e^{iu}$ are just the analogues of parallel wave fronts, the distance function $d(\omega;z,z_0)$
showing up in Eq.(\ref{wavedistance}) satisfies the addition rules exploited in the Huygens envelope construction\cite{BV}
\begin{subequations}\label{Huygens1}
\begin{align}
d(\omega;z,z_0)&=d(\omega;z,z_0^{\prime})+d(\omega;z_0^{\prime},z_0)\\
d(\omega;z_0^{\prime},z_0)&=-d(\omega;z_0,z_0^{\prime}).
\end{align}
\end{subequations}
Now $z=\tanh\frac{\varrho}{2}e^{i\varphi}\in{\mathbb D}$ corresponds to the fast variables $(\varrho,\varphi)$ on the other hand the pair $(z_0,z_0^{\prime})$ determines $(u,v)$ (though in an ambiguous manner) hence corresponds to the slow ones.
Eq.(\ref{Huygens1}) shows that the exponential 
\begin{equation}\label{BP}
e^{i\sigma_{\pm k}}\equiv e^{(1/2\pm ik)d(\omega;z_0, z_0^{\prime})}
\end{equation}
is depending merely on the slow variables.
Finally taking the exponential of Eq.(\ref{Huygens1}) in an obvious notation
we get
\begin{equation}\label{GT}
\psi^{\prime\pm}_k=\psi^{\pm}_ke^{i\sigma_{\pm k}}.
\end{equation}
In this equation we omitted the arguments. Clearly unlike the exponential of Eq.(\ref{BP}) the eigenstates $\psi_{\pm k}^{\prime}$ and $\psi_{\pm k}$ are depending on both type of variables.

Notice that the exponential of Eq.(\ref{BP}), which can be regarded as an analogue of the gauge degree of freedom familiar from Berry's Phase, is not a phase factor. It is rather an element of the group $GL(1,\mathbb C)$ of nonzero complex numbers.
This factor is coming from a change of reference point used for monitoring the plane waves after or before being scattered.
Unlike in the Euclidean case where for plane waves this change of reference results in merely a phase shift in the hyperbolic case the change alters the amplitude as well.

An alternative way of accounting for this gauge degree of freedom is as follows.
The geodesic operator of Eq.(\ref{hamiltoni}) is just the adjoint orbit of the one $\mathcal{H}_0=J_2$ which is a boost. In the coordinates of $\mathbb U$ it can be written as $J_2=i(\xi\partial_{\xi}+\eta\partial_{\eta})$. In the bulk it is generating $SO(1,1)$ transformations in the form $e^{i\beta J_2}$.  These encapsulate translations by $\beta$ along spacelike geodesics.
In the $AdS_3/CFT_2$ language it corresponds to the antisymmetric combination of conformal transformations inducing a flow from the left to the right endpoints of the causal diamond on the boundary\cite{Czech2}.
This transformation is the analogue of the symmetric combination related to the modular Hamiltonian which implements the flow from the bottom to the top of the causal diamond. 
Now since $e^{i\beta J_2}$ is trivially commuting with $\mathcal{H}_0$ such transformations are acting on the (\ref{base}) eigenfunctions as follows
\begin{equation}\label{horofazis}
\eta^{1/2\pm ik}\mapsto \left(\frac{\eta}{\Delta}\right)^{1/2\pm ik},\qquad \beta\equiv \log\Delta.
\end{equation}
Comparing this with Eq.(\ref{simplelambda2}) of our Appendix shows that the parameter $\Delta$ is related to the Euclidean diameter of the regularizing horocycle. 
Clearly Eq.(\ref{horofazis}) is of the (\ref{GT}) form for $\psi_{0k}^{\pm}$, linking the gauge degree of freedom to the freedom of choosing horocycles in an explicit manner. Moreover, due to the (\ref{coadjoint}) adjoint form this interpretation also carries through for an arbitrary eigenstate $\psi_k^{\pm}$.
Group theoretically the factor $e^{i\sigma_{\pm}}$ can be regarded as a representation of the $SO(1,1)$ boost transformations on our scattering states
in the form of a local $GL(1,\mathbb C)$ group element.
This replaces the usual $U(1)$ phase factor familiar from Berry's Phase.

The relationship found here between horocycles and the gauge degree of freedom for the scattering eigenstates of our geodesic operator has many virtues.
One of them is that it geometrizes nicely the boundary gauge freedom in the bulk.
As nicely summarized in Ref.\cite{Czech2} in a boundary theory without a scale, setting a cutoff is a gauge choice.
This implies that the space of gauge choices is a natural object of study. In this respect the somewhat strange notion: "the space of cutoffs" should be replaced by the mathematically well-defined one "the space of horocycles" $\mathbb G$ which is according to Appendix B. a homogeneous space just like our kinematic space $\mathbb K$.

\section{The Wigner Delay}

An arbitrary scattering state $\Psi_k(z)$ can be expanded with respect to a complete system of plane waves.
Such a system is formed by either incoming our outgoing sets of the form\cite{Helgason}  
\begin{subequations}
\begin{align}
\psi_{k}^{+}(z,u)&= e^{-iu J_3}\psi_{0k}^{+}(\varrho,\varphi; 0,\pi)\\
\psi_{k}^{-}(z,v)&= e^{-iv J_3}\psi_{0k}^{-}(\varrho,\varphi; 0,\pi).
\end{align}
\end{subequations}
These expressions are trivially related to the one of Eq.(\ref{helga}) where on the left hand side the fast variables are parametrized as $z=\tanh\frac{\varrho}{2}e^{i\varphi}$.
Then we have the expansions
\begin{subequations}\label{Wig}
\begin{align}
\Psi_k(z)&=\frac{1}{2\pi}\int_0^{2\pi}c_k^+(u)\psi_k^+(z,u)du\\
&= \frac{1}{2\pi}\int_0^{2\pi}c_k^-(v)\psi_k^-(z,v)dv.
\end{align}
\end{subequations}
According to the first expansion our wave function is a superposition of plane waves emitted from all boundary points $\omega=e^{iu}$, with $c_k^+(u)$ giving rise to a density distribution of these sources.
The second expansion encapsulates a superposition of plane waves absorbed at all boundary points $e^{iv}$ where $c_k^-(v)$
is the density distribution of sinks. These representations for $\Psi_k(z)$ were called in Ref.\cite{BV} as the past and future representations.

The density distributions of the sources and sinks can be related in the following manner\cite{BV,Kerimov}
\begin{equation}\label{relate2}
c_-(v)\equiv (\hat{S}c_+)(v)=\frac{1}{2\pi}\int_0^{2\pi}K(v,u)c_+(u)du
\end{equation}
where
\begin{equation}
K(v,u)=\frac{\sqrt{\pi}\Gamma(1/2+ik)}{\Gamma(-ik)}\vert\sin(v-u)/2\vert^{-1-2ik}
\label{kernelfv}
\end{equation}

Using the fact that the functions $\vert m\rangle =e^{imu}$ form an $SO(2)$ basis for $L^2(S^1)$ we get
\begin{equation}
\langle m^{\prime}\vert \hat{S}\vert m\rangle ={\delta}_{m^{\prime}m}S_m
\end{equation}
\noindent
with
\begin{equation}
S_m=\frac{\Gamma(1-ik)\Gamma(1/2+ik+m)}{\Gamma(1+ik)\Gamma(1/2-ik+m)}
\end{equation}
\noindent
$S_m$ turns out to be the scattering matrix of a one dimensional scattering problem in a P\"oschl-Teller potential
\begin{equation}
V(\varrho)=\frac{m^2-1/4}{\sinh^2\varrho}
\label{PT}
\end{equation}
\noindent
which shows up in the Hamiltonian $H=-\triangle$ taking care of the quantization of the geodesic motion.
This observation identifies the operator $\hat{S}$ as the scattering operator.
Indeed, according to Ref.\cite{Kerimov,Kerimov2} the fact that $\hat{S}$ acts as an intertwiner between Weyl equivalent principal series representations of $SO(2,1)$ enables this object to describe scattering situations in a purely group theoretical manner,
an approach which has already been taken up within the framework of algebraic scattering theory\cite{Iachello}.

Notice in particular when looking at the structure of $\hat{S}$ in the $SO(2)$ basis the $m=1/2$ case corresponds to the free case resulting in $S_{1/2}=1$ for the potential (\ref{PT}).
Moreover, the $m=0$ case for $S_0$ (up to a sign) gives the combination showing up in the phase of the kernel $K(u,v)$.
Moreover the kernel $K(v,u)$ is depending on the scattering energy $k\simeq\sqrt{E}$ and the quantity $v-u$ which is the length of the interval in the boundary. This rings a bell in connection with the Ryu-Takayanagi conjecture, where the length of a geodesic in the bulk turned out to be related to the entanglement entropy of a domain of length $v-u$ in the boundary.

In order to further elaborate on this point let us form the phase of the kernel in the form
\begin{equation}
S^2(k;v,u)\equiv\frac{K(v,u)}{K^{\ast}(v,u)}.
\end{equation}
\noindent
Defining formally a quantity reminiscent of the Wigner delay\footnote{The Wigner time delay is usually defined as the quantity $i\frac{d}{dE}\log Det S(E)$ where $E$ is the scattering energy and $S(E)$ is the scattering matrix. In our static case we rather considered an analogous quantity which is simply called Wigner delay in Ref.\cite{Kakukk}.}  as
\begin{equation}
d(k;v,u)\equiv i\frac{d}{dk}\log Det S(k;v,u)
\end{equation}
\noindent
we obtain
\begin{equation}
d(k;v,u)=\log\sin^2(v-u)/2-{\partial}_k\Delta(k)  
\end{equation}
where
\begin{equation}
e^{2i\Delta(k)}=\frac{\Gamma(1/2-ik)\Gamma(ik)}{\Gamma(1/2+ik)\Gamma(-ik)}.
\end{equation}
After some work the second term can be written in a form
\begin{equation*}
-{\partial}_k\Delta(k)=2\left(\Re{\psi(2ik)}-\Re{\psi(ik)}-\log 2\right)\equiv 2\Lambda(k)
\end{equation*}
Hence the final result can be written in the form
\begin{equation*}
d(k;v,u)=\log(e^{2\Lambda(k)}\sin^2(v-u)/2).
\end{equation*}
\noindent
Comparing this with the explicit expression for the (\ref{Suv}) entanglement entropy
we see that in our special case the delay is related to the holographic entanglement entropy with a k-dependent cutoff.

\section{Ptolemy relation, shear coordinates and strong subadditivity}

Let us choose for regularized length calculations of bulk geodesics non overlapping horocycles (see Figure 6.). Then due to the Ryu-Takayanagi relation for a boundary interval $A$ and the corresponding bulk geodesic ${\bf A}$ anchored to it one should have 
\begin{equation}
S_{A}=\frac{1}{2G_N}\log\lambda({\bf A})=\frac{c}{3}\log\lambda({\bf A}).
\label{RTlambda}
\end{equation}
Here we have used the Brown-Henneaux relation\cite{BH} $c=\frac{3R}{2G_N}$ with $R=1$, ${c}$ the central charge of the $CFT_2$, and the relationship between the lambda length and the regularized geodesic distance of Eq.(\ref{lambdalength}).

In the mathematics literature the lambda lengths serve as coordinates for the decorated Teichm\"uller space\cite{Penner}. They are positive. When we use their logarithm they can be related to entanglement entropy only for the physically meaningful case when the horocycles are tiny hence non overlaping, meaning that the lambda lengths are much greater then unity. According to Appendix A. the horocycles are Euclidean circles with radius inversely proportional to the third component (the height) of the corresponding light-like vector belonging to the future light cone ${\mathcal L}^+$. Hence in physical applications the tiny Euclidean circles used for regularization correspond to null vectors ${\mathcal L}^+$ with heights very large. Since ${\mathcal L}^+$ can be identified with the space of horocycles $\mathbb G$ this means that in this case we merely explore a portion of $\mathbb G$.  

However, in order to exploit the advantages of the full space of horocycles, correspondingly to use the full gauge degree of freedom for exploring new physical implications, it is rewarding to lift the restriction on the range of lambda lengths. This amounts to using the full space ${\mathcal L}^+$ for our  considerations. Thus in this most general case for overlapping horocycles lambda lengths can be smaller than unity, and for horocycles touching each other lambda lengths can even be equal to one.
As a result of this observation it is worth exploring the consequences of the more general formula 
\begin{equation}
S_{A}=\frac{c}{3}\vert \log\lambda({\bf A})\vert.
\label{RTlambda2}
\end{equation}
A byproduct of this generalization is the possibility of gauging away the entanglement entropies for any sequence of consequtive boundary intervals $I_1,I_2,\dots ,I_n$ giving rise to geodesic $n$-gons with geodesics ${\bf I}_1,{\bf I}_2,\dots,{\bf I}_n$ serving as their sides in the bulk.
We have something more to say on this possibility in the next section.

For the time being we use these observations as indications that the quantities one should really focus on are the gauge invariant ones.
First we turn our attention to a well-known quantity of that kind.

Let us denote subregions of the boundary $\partial\mathbb D$ by the letters $A,B,C,D,E,F$, the corresponding geodesics of the bulk $\mathbb D$ anchored to them by ${\bf A,B,C,D,E,F}$, and the corresponding points in kinematic space $\mathbb K$ by ${\mathcal A},{\mathcal B},{\mathcal C},{\mathcal D},{\mathcal E},{\mathcal F}$.
According to strong subadditivity for regions $E$ and $F$ on the boundary $\partial{\mathbb D}$ for the von-Neumann entropies one has
\begin{equation}
S_{E}+S_{F}\geq S_{{E}\cup{{F}}}+S_{{E}\cap{F}}.
\label{SSA}
\end{equation}
We choose a quadrangle as shown in Figure 4. with
\begin{gather}
E=A\cup B,\qquad F=B\cup C\\ D=A\cup B\cup C=E\cup F,\qquad E\cap F=B.
\label{choice}
\end{gather}
According to the Ptolemy relation proved in Refs.\cite{Penner,Pennerbook}
we have
\begin{equation}
\lambda(\bf A)\lambda(\bf C)+\lambda(\bf B)\lambda(\bf D)=\lambda(\bf E)\lambda(\bf F).
\label{ptolemy}
\end{equation}
\begin{figure}
{\includegraphics[width=\columnwidth]{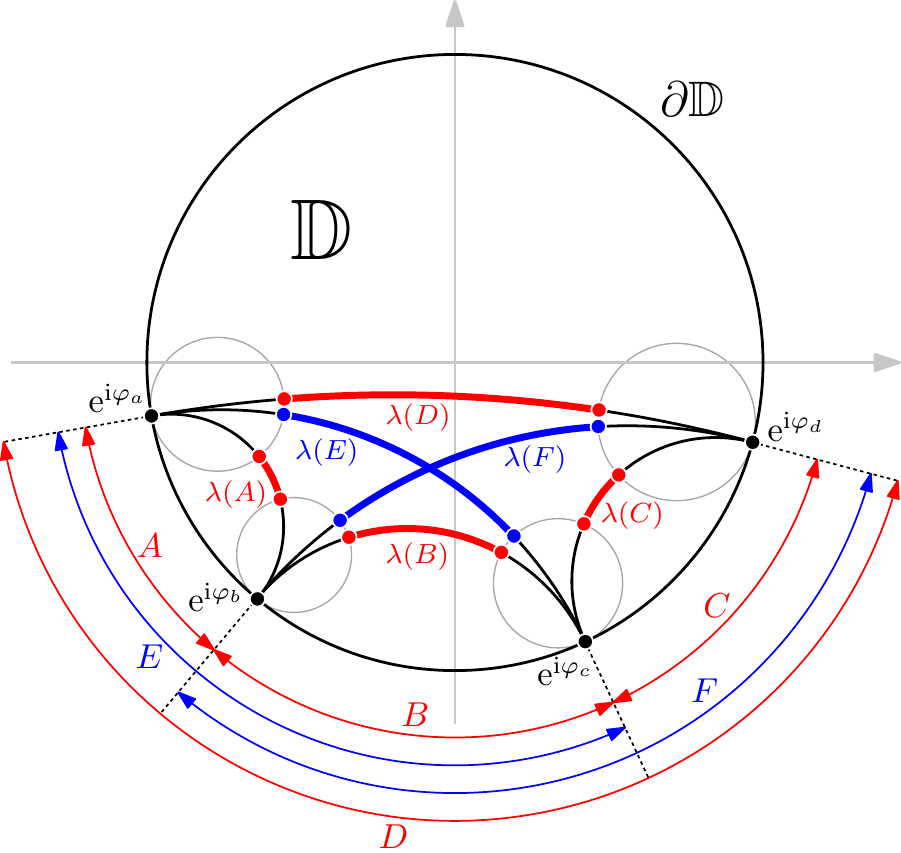}}
\caption{A quadrangle in ${\mathbb D}$ used for considerations connected to the Ptolemy relation Eq.(\ref{ptolemy}).}
\end{figure}
Let us write this in the form
\begin{equation}
1+\frac{\lambda(\bf A)\lambda(\bf C)}{\lambda(\bf B)\lambda(\bf D)}=
\frac{\lambda(\bf E)\lambda(\bf F)}{\lambda(\bf B)\lambda(\bf D)}.
\label{Ptolemy2}
\end{equation}
Taking the logarithm and using Eqs.(\ref{RTlambda})-(\ref{choice}) then yields
\begin{equation}
S_E+S_F-S_{E\cup F}-S_{E\cap F}=\frac{c}{3}\log\left(1+\frac{\lambda(\bf A)\lambda(\bf C)}{\lambda(\bf B)\lambda(\bf D)}\right)\geq 0
\label{ssshear}
\end{equation}

One can easily relate this result to a well-known one expressing strong subadditivity in terms of the cross ratio\cite{CS}.
Indeed, notice that by virtue of Eqs.(\ref{ze})-(\ref{tau}) we have
\begin{equation*}
[d,b;a,c]\equiv
\frac{(\xi_d-\xi_a)(\xi_b-\xi_c)}{(\xi_d-\xi_c)(\xi_b-\xi_a)}=
\frac{\sin\left(\frac{\varphi_d-\varphi_a}{2}\right)
\sin\left(\frac{\varphi_b-\varphi_c}{2}\right)
}
{\sin\left(\frac{\varphi_d-\varphi_c}{2}\right)
\sin\left(\frac{\varphi_b-\varphi_a}{2}\right)}
\label{crossratio}
\end{equation*}
where $[d,b;a,c]$ is the cross ratio
on ${\mathbb R}{\mathbb P}^1$.
Let us denote the negative of this cross ratio by $t(\bf E)$, then using Eq.(\ref{simplelambda}) one gets the alternative expression
\begin{equation}
t({\bf E})=-[d,b;a,c]=\frac{\lambda({\bf B})\lambda({\bf D})}{\lambda({\bf A})\lambda({\bf C})}.
\label{te}
\end{equation}
Hence $t(\bf E)$ has the dual interpretations as the ratio of Euclidean lengths $\frac{BD}{AC}$ in the boundary $\partial{\mathbb U}$ of the upper half plane, and also as the ratio
of lambda lengths $\frac{\lambda(\bf B)\lambda(\bf D)}{\lambda(\bf A)\lambda(\bf C)}$ in the bulk $\mathbb U$.

Notice now that unlike a lambda length $\lambda(\bf E)$ the quantity $t(\bf E)$ is not depending on the horocycles, it is regularization independent. Moreover, one can prove\cite{Pennerbook} that one can regard the quantities $t({\bf E})$ and $t({\bf F})$ with the latter defined by 
\begin{equation}
t({\bf F})=-[b,d;a,c]=\frac{\lambda({\bf A})\lambda({\bf C})}{\lambda({\bf B})\lambda({\bf D})}
\label{Teichm}
\end{equation}
as coordinates associated to the diagonal edges showing up in figure 4.  More precisely, $t({\bf E})$ and $t({\bf F})$ serve as alternative local coordinates for the Teichm\"uller space, which is the space of shapes of bulk geodesic quadrangles.

The meaning of $t({\bf E})$ is as follows (see Corollary 4.16. of Ref.\cite{Pennerbook}). Take a geodesic starting from $e^{i\varphi_d}$ and reaching the diagonal ${\bf E}$ in a right angle. Similarly take another geodesic from $e^{i\varphi_b}$ and reaching the diagonal ${\bf E}$ from the other side again in a right angle. These geodesics do not generally intersect in the same point. Let us call these points midpoints. If in the upper half plane the point $\xi_a$ is sent to zero and $\tau_c$ to $i\infty$, then the distance between these midpoints is the shear $\log(\vert\xi_d/\xi_b\vert)\geq 0$ if $\vert\xi_d\vert\geq \vert\xi_b\vert$. Hence $\log t({\bf E})$ and  $\log t({\bf F})$ have  the meaning as a shears associated to the diagonal edges ${\bf E}$ and ${\bf F}$. 
The illustration of the meaning of this shear coordinate can be found in Figure 5.
\begin{figure}
{\includegraphics[width=\columnwidth]{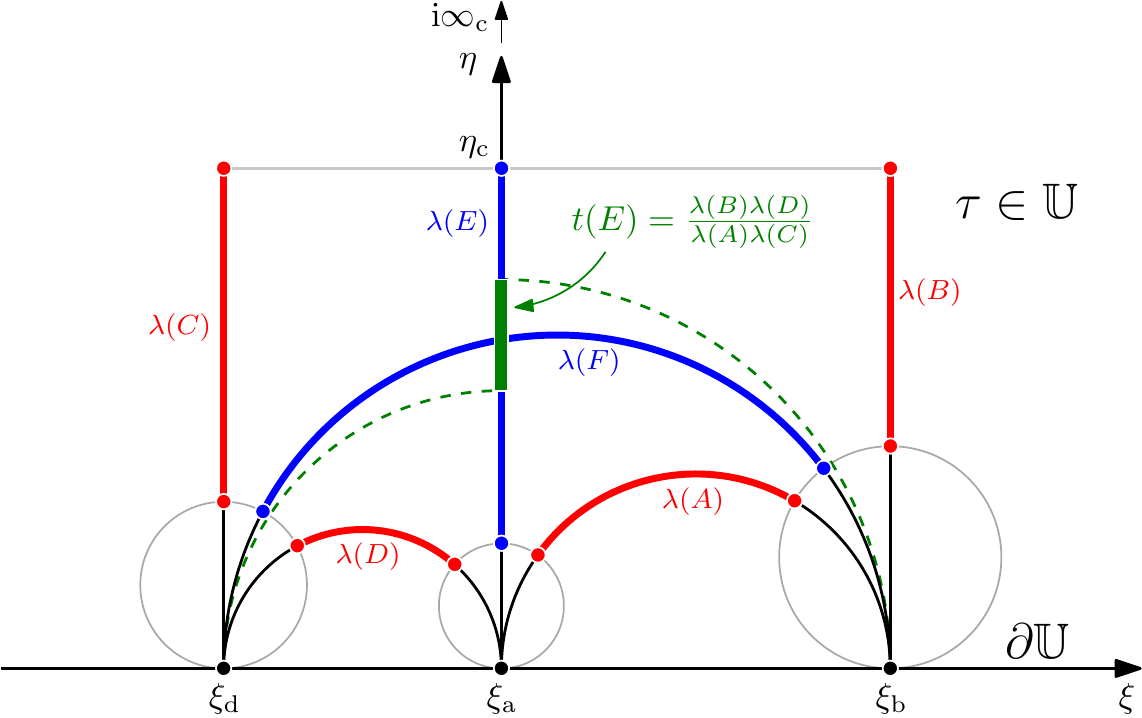}}
\caption{The quadrangle of Figure 4. depicted now in the upper half plane ${\mathbb U}$. The meaning of the shear coordinate $t({\bf E})$ is explained.}
\end{figure}

A canonical arrangement in $\mathbb D$ can be obtained when $(e^{i\varphi_a},e^{i\varphi_b},e^{i\varphi_c},e^{i\varphi_d})=(-1,-i,1,e^{i\varphi_d})$  with $\frac{\pi}{2}<\varphi_d<\pi$. On the other hand modifying Figure 5. accordingly, in $\mathbb U$ this arrangement corresponds to sending the triple of boundary points $(a,b,c)$ to $(\xi_a,\xi_b,\xi_c)=(0,1,\infty)$, and the fourth point $d$ is located in the open interval $\xi_d\in (-1,0)$. Then $\log t(E)$ is negative and $\log t(F)$ is positive. Reinterpreting then in the canonical arrangement the shear $\log t(F)$ in terms of the von Neumann entropies we get
\begin{equation}
\frac{c}{3}\log t(F)= S(A)+S(C)-S(B)-S(D)> 0.
\label{shearentropy}
\end{equation} 
Clearly for $\xi_d\leq -1$ one has $\log t(F)\leq 0$.

We emphasize that the shear is independent of regularization.
Since regularization amounts to using horocycles, and the degree of freedom of choosing different horocycles corresponds in the Berry's Phase language to choosing different local phase factors for our scattering states, one can say that the sheer coordinates are gauge invariant ones.
Moreover, since for pure states $S(D)=S(\overline{D})$ the gauge invariant shear $\log t(F)$ provides the bulk dual of the boundary quantity  $S(A)+S(C)-S(B)-S(\overline{D})$ characterizing any fourpartite splitting of the boundary ${\mathbb R}{\mathbb P}^1$ into disjoint intervals of the form: ${\mathbb R}{\mathbb P}^1=A\cup B\cup C\cup\overline{D}$.
Hence we managed to find a relationship between a boundary related entropic quantity characterizing fourpartite splits, and the bulk related shear coordinate characterizing the Teichm\"uller space of quadrangles.

We can summarize our results in the form
\begin{subequations}\label{ssshear2}
\begin{align}
S_E+S_F-S_{E\cup F}-S_{E\cap F}=\frac{c}{3}\log\left(1+t({\bf F})\right)\geq 0\\
S_{E}+S_{\overline{F}}-S_{E\cup \overline{F}}-S_{E\cap \overline{F}}=\frac{c}{3}\log\left(1+t({\bf E})\right)\geq 0
\end{align}
\end{subequations}
where in the first of these equations the overlap is given by region $B$ and in the second region $A$. 
As explained in \cite{Czech1} the left hand sides of these measures of strong subadditivity can be expressed in terms of the conditional mutual informations as
\begin{subequations}\label{ssshear4}
\begin{align}
I(A,C\vert B)=\frac{c}{3}\log\left(1+t({\bf F})\right)\geq 0\\
I(B,D\vert A)=\frac{c}{3}\log\left(1+t({\bf E})\right)\geq 0
\end{align}
\end{subequations}
where
\begin{equation}
I(A, C\vert B)\equiv S(A \vert B)-S(A\vert BC)
\end{equation}
with $S(A\vert B)=S(AB)-S(B)$ is the conditional entropy. 
The quantities on the left hand sides indicate that conditioning on a larger subsystem can only reduce the uncertainty about a system.

Let us now use
\begin{equation}
t({\bf F})=\frac{\lambda({\bf A})\lambda({\bf C})}{\lambda({\bf B})\lambda({\bf D})}
=
\frac{\vert\sin\left(\frac{\varphi_b-\varphi_a}{2}\right)\vert
\vert\sin\left(\frac{\varphi_d-\varphi_c}{2}\right)\vert
}
{\vert\sin\left(\frac{\varphi_d-\varphi_a}{2}\right)\vert
\vert\sin\left(\frac{\varphi_c-\varphi_b}{2}\right)\vert}
\label{crossratio2}
\end{equation}
with the special choice of Ref.\cite{Czech1}
\begin{equation}
(\varphi_a,\varphi_b,\varphi_c,\varphi_d)=(u-du,u,v,v+dv).
\label{choice2}
\end{equation}
In lowest order we obtain
\begin{equation}
t({\bf F})=\frac{dudv}{4\sin^2\left(\frac{v-u}{2}\right)}+\dots
\label{Crofi}
\end{equation}
Since in this case $\log(1+t({\bf F}))=t({\bf F})+\dots$ we can relate the result of Ref.\cite{Czech1}
to an infinitesimal Teichm\"uller coordinate as follows
\begin{equation}
S_E+S_F-S_{E\cup F}-S_{E\cap F}\simeq\frac{c}{3}t({\bf F})=\frac{{\partial}^2 S(u,v)}{{\partial}u{\partial}v}dudv \geq 0,
\label{ssshear3}
\end{equation}
where $S(u,v)$ is given by Eq.(\ref{Suv}).

There is yet another way of rewriting Eqs.(\ref{ssshear2}).
Indeed, according to Corollary 4.16.d of Ref.\cite{Pennerbook} one can write
\begin{equation}
1+t(F)=\cosh^2\frac{\ell}{2},\qquad
1+t(E)=\cosh^2\frac{\ell^{\prime}}{2}
\label{otherdist}
\end{equation}
where $\ell$ ($\ell^{\prime}$) is the infimum of the geodesic distances between the geodesics ${\bf B}$ and ${\bf D}$ (${\bf A}$ and ${\bf C}$).
Finally one can write
\begin{subequations}\label{important}
\begin{align}
S_E+S_F-S_{E\cup F}-S_{E\cap F}=\frac{1}{G_N}\log\cosh\left(\frac{\ell}{2}\right)\\
S_E+S_F-S_{E\cup \overline{F}}-S_{E\cap \overline{F}}=\frac{1}{G_N}\log\cosh\left(\frac{{\ell}^{\prime}}{2}\right)
\end{align}
\end{subequations}

Notice that the left hand sides of these equations
can be regarded as aggregate measures of how far the infrared degrees of freedom are away from saturating the strong subadditivity of entanglement.
As we see for a particular choice of subregions $E$ and $F$ we have a corresponding geodesic quadrangle. Depending on which side of the quadrangle we regard as a one arising from the intersection of regions related to $E$ and $F$ we have four possibilities. However, since for pure states $S(\overline{E})=S(E)$ etc. as far as entropic relations associated to intersections are concerned we have merely two possibilities. In the first case we have $E\cap F=B$ and in the second $E\cap\overline{F}=A$. These cases correspond to two possible triangulations of our quadrangle. In this respect in the first case the geodesics $\bf F$ and in the second $\bf E$ play a distinguished role with their sheer coordinates $t({\bf F})$ and $t({\bf E})$, satisfying $t({\bf F})t({\bf E})=1$, are quantifying our measures of conditional mutual information. 
Hence the bulk meaning $I(A,C\vert B)$ and $I(B,D\vert A)$ is connected to triangulations of geodesic quadrangles.
On the other hand we also know that the kinematic space meaning of these quantities are areas. 
Note in this respect that one can easily check that the chain rule for conditional mutual information of Ref.\cite{Czech1}, expressing the fact that areas in kinematic space are additive, is just another manifestation of the Ptolemy relation (\ref{ptolemy}) in the bulk. We will see later that this chain rule for conditional mutual information has its roots in the patching condition for local coordinates of the Teichm\"uller space for bulk geodesic polygons.

In order to approach these issues first we look at the kinematic space interpretation of our results.
Let us consider the canonically arranged quadrangle with $(\varphi_a,\varphi_b,\varphi_c,\varphi_d)=(\pi,3\pi/2,0,\varphi_d)$ with $\pi/2<\varphi_d<\pi$ and the following set of vectors in $\mathbb K$
\begin{equation*}
\mathcal{A}=\begin{pmatrix}-1\\-1\\1\end{pmatrix},\quad
\mathcal{B}=\begin{pmatrix}1\\-1\\1\end{pmatrix},\quad
\mathcal{C}=\begin{pmatrix}t\\1\\t\end{pmatrix}
\end{equation*}
\begin{equation*}
\mathcal{D}=\begin{pmatrix}-\frac{1}{t}\\1\\\frac{1}{t}\end{pmatrix},\quad
\mathcal{E}=\begin{pmatrix}0\\-1\\0\end{pmatrix},\quad
\mathcal{F}=\begin{pmatrix}1\\\frac{1-t}{1+t}\\\frac{t-1}{1+t}\end{pmatrix}
\end{equation*}
where
$0<t=\cot\frac{\varphi_d}{2}<1$ with $t=t({\bf E})$. 
According to Eq.(\ref{haha}) one can check that these vectors correpond to the set $({\bf A},\dots,{\bf F})$ of counter clock-wise {\it oriented geodesics}.
We remark that the quantity $\cos\vartheta\equiv\frac{1-t}{1+t}$ is related to the angle $\vartheta$ between the two diagonals of the quadrangle opposite to ${\bf D}$, yielding the relation $\cos^2\frac{\vartheta}{2}\cosh^2\frac{\ell^{\prime}}{2}=1$. This observation provides yet another interpretation of the measure of subadditivity in (\ref{important}) in terms of the angle $\vartheta$.

In the following the letters $(\overline{\mathcal A},\dots,\overline{\mathcal F})$ will denote the negatives of the corresponding vectors representing the oppositely oriented geodesics in $\mathbb K$.
Now using (\ref{otherdist}) we have
\begin{equation}
\cosh \ell=\overline{\mathcal{B}}\cdot\mathcal{D},\qquad
\cosh \ell^{\prime}=\overline{\mathcal{A}}\cdot\mathcal{C} 
\label{kincausal}
\end{equation}
expressing the infimum of the geodesic distances between the pairs of geodesics $({\bf B},{\bf D})$ and $({\bf A},{\bf C})$ in $\mathbb D$ in terms of the data of the corresponding pair of points $(\overline{\mathcal{B}},\mathcal{D})$ and
$(\overline{\mathcal{A}},\mathcal{C})$ in $\mathbb K$.
These results can be used to to express our boundary measures of (\ref{important}) of strong subadditivity in terms of kinematic space data.

One can check that the vectors $\mathcal{D}-\overline{\mathcal{B}}$, and
$\mathcal{C}-\overline{\mathcal{A}}$
 are future directed time-like ones in ${\mathbb R}^{2,1}$.
Moreover, one can consider the duad of quadruplets of points 
$(\mathcal{D},\overline{\mathcal{F}},\overline{\mathcal{B}},\overline{\mathcal{E}})$ and
$(\mathcal{C},{\mathcal{F}},\overline{\mathcal{A}},\overline{\mathcal{E}})$.
One can then verify that these quadruplets form a causal diamond-like structure.
Indeed, for example in the first case $\overline{\mathcal{F}},\overline{\mathcal{E}}$ are space-like separated, and 
the vectors $\mathcal{D}-\overline{\mathcal{F}}$ etc. are future directed and null.
This structure is in accord with the observation of Ref.\cite{Czech1} that boundary intervals are organized according to causal structures formed by points located in kinematic space.
These structures date back to the natural causal structure bulk geodesics enjoy based on the containment relation of boundary intervals. Geodesics are time-like separated if they have embedded boundary intervals, null separated if they share a left or right endpoint, and space-like separated if their boundary intervals are not embedded.
Notice in this respect that Eq.(\ref{ssshear3}) familiar from Ref.\cite{Czech1} is just the infinitesimal version of our general formula of Eq.(\ref{important}).

As a simple application of Eqs.(\ref{otherdist})-(\ref{kincausal}) one can consider the kinematic space of the static BTZ black hole configuration\cite{Zhang}. In this case one starts with a Fuchsian group $\Gamma=\{\gamma^n\vert \gamma\in PSL(2,\mathbb R), n\in\mathbb Z\}$ where $\gamma$ is the fundamental element responsible for the identification of the geodesics $\bf B$ and ${\bf D}$ of Figure. 4. yielding the BTZ black hole configuration with boundary components $A$ and $C$. Then the horizon length of the BTZ black hole ${\ell}$ is given by the formula\cite{Maxfield,Zhang} 
$\vert {\rm Tr}\gamma\vert =2\cosh\frac{\ell}{2}$. Hence we see that formally the horizon length is related to the shear coordinate $t(\bf F)$ via Eq.(\ref{otherdist})
and the corresponding conditional mutual information $I(A,C\vert B)$.
Notice also that the proper time $\Delta\tau$ between the two points $\overline{\mathcal B}$ and ${\mathcal D}$ along a timelike geodesic in $\mathbb K$ is just the geodesic length $\ell$ between the geodesics ${\bf B}$ and {\bf D} i.e. we have\cite{Zhang} $\Delta\tau=\ell$.

Moreover back to the pure static $AdS_3$ case, according to Figure 6. of 
Ref.\cite{Zhang} we have four causal diamonds showing up in two-fold degenerate pairs corresponding to the independent conditional mutual informations of $I(A,C\vert B)$ and $I(B,D\vert A)$  located at the central belt of $\mathbb K$ partitioned into $20$ domains. They are defined by the quadruplets of points   
$(\mathcal{D},\overline{\mathcal{F}},\overline{\mathcal{B}},\overline{\mathcal{E}})$ and
$(\mathcal{C},{\mathcal{F}},\overline{\mathcal{A}},\overline{\mathcal{E}})$ taken together with the quadruplets replacing the corresponding vectors by their overlined versions. Notice in this respect that the transformation $t({\bf F})\mapsto 1/t({\bf F})$
results in a flip between $I(A,C\vert B)$ and $I(B,D\vert A)$. We will see in the next section that this is the fundamental transformation of the $A_1$ cluster algebra. In kinematic space then a flip of $A_1$ type yields a change between the correponding causal diamonds having a common point ($\overline{\mathcal E})$, and at the same time recording also the change in the corresponding areas i.e. conditional mutual informations

\begin{equation}
I(A,C\vert B)-I(B,D\vert A)=\frac{c}{3}\log t({\bf F}).
\label{shearRT}
\end{equation}
Hence symbolically we have the rule: a change in area under a flip equals $\frac{c}{3}$ times shear.
Note that a $\delta A =\frac{c}{3}\log t$ type relation is of the (\ref{RTlambda}) form.

Now for the BTZ black hole we have to chose a fundamental domain in $\mathbb K$. An example for such a domain can be seen in Figure 8. of Ref.\cite{Zhang}. In this case we have merely one of the independent causal diamonds at the central belt at our disposal, i.e. the one belonging to our fundamental domain. As a result the previously mentioned flip is not a legitimate transformation. This is an indication of the fact that as we change the quantum state on the boundary  the corresponding cluster algebraic structure of the bulk should change accordingly.   

In order to further elaborate on this observation rather than a quadrangle one should consider a geodesic polygon.
First of all let us notice that for a boundary interval $X$ the quantities $\log\lambda({\bf X})$ (regularization dependent lambda lengths, related to boundary entropies) and $\log t({\bf X})$ (regularization independent shear coordinates, related to conditional mutual informations) provide coordinates for the decorated Teichm\"uller space, and Teichm\"uller space respectively of such polygons\cite{Penner,Pennerbook}.
Moreover, unlike $\log\lambda({\bf X})$ the shear $\log t({\bf X})$ is depending on the context provided by some quadrangle for which ${\bf X}$ shows up as a diagonal for one of its two possible triangulations. 
This implies that when changing a triangulation a change in a shear coordinate of an edge also depends on how its neighbors change. 

To see this consider a pentagon, arising from the quadrangle of Figure 4. by adjoining an extra point $a^{\prime}$ in the boundary with angular coordinates $\varphi_a<\varphi_{a^{\prime}}<\varphi_b$. As a result of this the boundary interval $A$ is now divided into two sub-intervals: $A_1$ and $A_2$.
Then if we change the triangulation of the {\it original} quadrangle ${\bf ABCD}$ from a one featuring ${\bf E}$ as its diagonal to the dual one featuring ${\bf F}$, then the status of ${\bf A}$ is changing from a diagonal belonging to quadrangle ${\bf A_1A_2BE}$ to a one belonging to ${\bf A_1A_2FD}$. 
As a result of this 
by embedding the remaining sides inside of further pentagons
and
using the analogues of Eqs.(\ref{te})-(\ref{Teichm}) one obtains the transformation rules
\begin{subequations}
\begin{align}
\log t({\bf A})\mapsto \log t({\bf A})- \log\left({1+t({\bf F})}\right)\\
\log t({\bf C})\mapsto \log t({\bf C})- \log\left({1+t({\bf F})}\right)\\
\log t({\bf B})\mapsto \log t({\bf B})+ \log\left({1+t({\bf E})}\right)\\
\log t({\bf D})\mapsto \log t({\bf D})+ \log\left({1+t({\bf E})}\right)
\label{egy1}
\end{align}
\end{subequations}
These rules show that the shear coordinates change by the basic measures of strong subadditivity of Eqs.(\ref{important}), i.e. by the conditional mutual informations of Eqs.(\ref{ssshear4}).
Since the shear coordinates are local coordinates for the space of shapes i.e. Teichm\"uller space\cite{Pennerbook} for geodesic polygons, and the (66) equations are reminiscent of patching conditions for local coordinates this result gives an interesting physical interpretation for conditional mutual information.
In order to explore further the implications of this result in the next section we formulate our considerations in terms of the language provided by cluster algebras.

\section{Cluster algebras}

The basic role our conditional mutual informations are playing in our considerations of geodesic $n$-gons dates back to the fundamental one of the Ptolemy relation of Eq.(\ref{ptolemy}) playing in the generation of an algebraic structure called the $A_{n-3}$ cluster algebra\cite{Williams}.

In order to see this let us turn back to parametrizing the sides of a triangulation of an $n$-gon by lambda lengths.
For a triangulation of an $n$-gon we have $2n-3$ edges, $n-3$ of them are diagonal ones. 
For example in the $n=5$ case one can adjoin an extra boundary point $e^{i\varphi_{d^{\prime}}}\in{\mathbb D}$ with $\varphi_{d^{\prime}}< \varphi_d$, to the canonically arranged quadrangle 
 $(e^{i\varphi_a},e^{i\varphi_b},e^{i\varphi_c},e^{i\varphi_d})=(-1,-i,1,e^{i\varphi_d})$  with $\frac{\pi}{2}<\varphi_d<\pi$.
The five boundary points with horocycles can be choosen by  fixing the corresponding null vectors as
\begin{equation*}
{\bf a}=\begin{pmatrix}-1\\0\\1\end{pmatrix},\quad
{\bf b}=\begin{pmatrix}0\\-1\\1\end{pmatrix},\quad
{\bf c}=\begin{pmatrix}1\\0\\1\end{pmatrix}
\end{equation*}

\begin{equation*}
{\bf d}=\frac{1}{1+\cos\varphi_d}\begin{pmatrix}\cos\varphi_d\\ \sin\varphi_d\\ 1\end{pmatrix},\quad
{\bf d^{\prime}}=\frac{1}{1-\cos\varphi_{d^{\prime}}}\begin{pmatrix}\cos\varphi_{d^{\prime}}\\ \sin\varphi_{d^{\prime}}\\ 1\end{pmatrix}
\end{equation*}
 
In this case 
${\bf a}\cdot{\bf b}={\bf b}\cdot{\bf c}={\bf c}\cdot{\bf d}^{\prime}={\bf d}^{\prime}\cdot{\bf d}={\bf d}\cdot{\bf a}=-1$ provided 
\begin{equation*}
\cot\frac{\varphi_{d^{\prime}}}{2}=(1+\sqrt{2})\cot\frac{\varphi_{d}}{2}.
\end{equation*}
This gives us an example of the consequtive horocycles touching each other hence for the lambda lengths we have $\lambda_{ab}=\lambda_{bc}=\dots=\lambda_{da}=1$. This amounts to using the gauge degree of freedom to scale all of the entanglement entropies $S_{ab}=S_{bc}=\dots =S_{da}$ to zero.
Clearly this trick can be done for arbitrary choice of pentagons (or $n$-gons) not merely the canonical one, provided we use the full gauge degree of freedom provided by our space of horocycles.

Though we can scale away the regularized entropies of the sides of our pentagon  we cannot do this simultaneously for the diagonals. We have five possible choices for the diagonals, any two of them starting from the same vertex resulting in a triangulation for the pentagon. We have five such triangulations and one can choose the one based on the diagonals $ac$ and $cd$ as a basic one. For this basic triangulation the corresponding null vectors then give
$-{\bf a}\cdot{\bf c}=2$ and $-{\bf c}\cdot{\bf d}=\tan^2\frac{\varphi_d}{2}>1$.
Using (\ref{lambdalength}) one can define their lambda lengths
\begin{subequations}\label{clustervar}
\begin{align}
x_1\equiv \lambda({\bf a},{\bf c})=\sqrt{2},\\
x_2\equiv \lambda({\bf c},{\bf d})=\tan\frac{\varphi_d}{2}
\end{align}
\end{subequations}
hence the corresponding regularized geodesic lengths are
\begin{subequations}\label{clusterentr}
\begin{align}
d({\bf a},{\bf c})=\log 2>0,\\
d({\bf c},{\bf d})=2\log\tan\frac{\varphi_d}{2}>0
\end{align}
\end{subequations}
then the associated (regularized) entanglement entropies for these diagonals are nonzero.

For the remaining diagonals one gets
\begin{subequations}\label{clustervar2}
\begin{align}
x_3\equiv \lambda({\bf b},{\bf d})=\frac{1}{\sqrt{2}}(1+\tan\frac{\varphi_d}{2})\\
x_4\equiv \lambda({\bf b},{\bf d}^{\prime})=\frac{1}{\sqrt{2}}(1+(1+\sqrt{2})\cot\frac{\varphi_{d}}{2})\\
x_5\equiv \lambda({\bf a},{\bf d}^{\prime})=(1+\sqrt{2})\cot\frac{\varphi_{d}}{2}.
\end{align}
\end{subequations}
Notice that two from the five lambda lengths are not necessarily greater than one, hence for the interpretation of their logarithms as entanglement entropies one should use the generalized formula of Eq.(\ref{RTlambda2}).

Now the important observation we would like to make is that the lambda lengths $x_i, i=1,\dots 5$ regarded as cluster variables satisfy the recursion relation
\begin{equation}
x_{i+1}x_{i-1}=x_i+1
\label{clusteralg}
\end{equation} 
where this relation should be understood modulo $5$.
As a result of this the (\ref{RTlambda2}) entanglement entropies are recursively related. Clearly this recursion relation is based on the (\ref{ptolemy}) Ptolemy relation. Indeed if we consider the basic triangulation $T$ and the quadrangle $abcd$ with diagonal having lambda length $x_1$, then there is a new triangulation $T^{\prime}$ which is obtained by replacing the diagonal of the quadrangle with its  other one. Then the Ptolemy relation directly gives $x_3x_1=x_2+1$. This local move is called a {\it flip}\cite{Williams}.
Proceeding in this manner in the next step one can use the flip operation for the quadrangle $bcd^{\prime}d$ with diagonal $cd$ having lambda length $x_2$ yielding the Ptolemy relation $x_4x_2=x_3+1$ etc. 
It is well-known that the resulting algebraic structure is called the $A_2$ cluster algebra. This algebraic structure is independent of the triangluation and our very special  arrangement of the pentagon.

The algebraic structure exemplified by the (\ref{clusteralg}) relation we have found here is a general pattern showing up for $n$-gons with $n\geq 4$, and is called the $A_{n-3}$ cluster algebra.    
According to the general theory of cluster algebras\cite{Williams} in our special case to the diagonals of the $n$-gon we associate {\it cluster variables} and to the sides {\it coefficient variables}. The first set of variables are related to nonzero regularized entanglement entropies, and the second set comprises the variables answering the entropies which are scaled away.  

We also note that one can identify the $A_{n-3}$ cluster algebra with the coordinate ring of the Grassmannian $Gr(2,n)$ of two-planes in an $n$-dimensional vector space $V$. In this context the set of Ptolemy relations for subquadrangles  correspond to the Pl\"ucker relations, providing sufficient and necessary conditions for an arbitrary bivector $P=\sum_{i<j}P_{ij}e_i\wedge e_j$ to be the separable i.e. of the $P=\alpha\wedge \beta$ form for $\alpha,\beta\in V$. 
For $n=4$ this relation is of the form: $P_{12}P_{34}-P_{13}P_{24}+P_{14}P_{23}=0$ in accord with the form of the (\ref{ptolemy}) relation. On this point see some more details in Appendix A.

Having found an algebraic structure in the bulk encapsulating the regularized entanglement patterns of the boundary the question arises whether there is a similar structure for gauge (regularization) independent quantities?
Based on our observations we have made in connection with Eqs.(66) the answer to this question is obviously yes. Indeed, shears like $\log t({\bf E})$ of Eq.(\ref{te}) appearing as logarithms of cross ratios are of that kind. 

In order to elucidate this algebraic structure again we take our canonically arranged pentagon with sides ${\bf A},{\bf B},{\bf C}^{\prime},{\bf D}^{\prime}, {\bf D}$ arranged in a counter clock-wise manner.
In this picture the diagonals  ${\bf E}$ and ${\bf C}$ both of them starting from the point $z=1\in\partial{\mathbb D}$ provide the basic triangulation of our pentagon. ${\bf E}$ is the diametrical geodesic of $\mathbb D$ serving as a diagonal of the quadrangle ${\bf ABCD}$, on the other hand ${\bf C}$ is the diagonal of the quadrangle ${\bf EC}^{\prime}{\bf D}^{\prime}{\bf D}$.  
Let us denote the corresponding shear coordinates as
\begin{equation}
u_1=t({\bf E}),\qquad u_2=t({\bf C})=\frac{\lambda({\bf E})\lambda({\bf D}^{\prime})}{\lambda({\bf C}^{\prime})\lambda({\bf D}) }.
\label{u1u2}
\end{equation}
Then applying a flip to the quadrangle ${\bf ABCD}$ turns the diagonal ${\bf E}$ to diagonal ${\bf F}$. Under this process the other diagonal ${\bf C}$ stays the same but will be featuring the new quadrangle ${\bf FBC}^{\prime}{\bf D}^{\prime}$. Proceeding then in the same manner as we did in the previous section resulting in Eqs.(66) we obtain the transformation
\begin{equation*}
\label{shetrans1}
(u_1,u_2)\mapsto (\frac{1}{u_1}, \frac{u_1u_2}{1+u_1})\mapsto
\end{equation*}
In the next step we take a flip to the quadrangle ${\bf FBC}^{\prime}{\bf D}^{\prime}$ changing ${\bf C}$ to a new diagonal say ${\bf G}$ and leaving ${\bf F}$ alone although in a new role, namely as a diagonal of a new quadrangle ${\bf AGD}^{\prime}{\bf D}$. Proceeding recursively after a period of five flips we get back to the initial configuration. The new terms of this set of transformations
provides the cross ratios  
\begin{equation*}
(\frac{u_2}{1+u_1+u_1u_2},\frac{1+u_1}{u_1u_2})\mapsto (\frac{1+u_1+u_1u_2}{u_2},\frac{1}{u_1(1+u_2)})\mapsto
\end{equation*}
\begin{equation*}
\label{shetrans2}
\mapsto(\frac{1}{u_2},u_1(1+u_2)) \mapsto (u_2,u_1).
\end{equation*}
Notice that after this sequence of five flips we get back to the initial set of cross ratios up to a permutation.
Hence the process really closes after a period of ten flips.

Our simple example can of course be generalized for $n-3$ cross ratios giving rise to $n-3$ sheer coordinates for geodesic $n$-gons ($n\geq 4$). In this case one settles with an algebraic structure, based on triangulations of such $n$-gons, with period $2n$.
The steps of transformations can then be neatly summarized in a system of recursion relations which is known in the literature as Zamolodchikov's $Y$-system of $A_{n-3}$ type. As it is known the solutions of this system of recursion relations is periodic of $2n$. This result was originally formulated as a conjecture by Zamolodchikov and proved by cluster algebra methods later\cite{Zamo,Williams}.   

Hence our observations on gauge invariant cross-ratios can also be formulated within the framework of cluster algebras.
Notice also that in our first encounter with cluster algebras for regularized entanglement entropies lambda lengths, in our second encounter for gauge invariant conditional mutual informations cross ratios of lambda lengths played a key role.
In the cluster algebra language in the first case entanglement entropies were associated to cluster variables of the diagonal geodesics of polygons, on the other hand the entropies associated to coefficient variables of the sides were scaled away. Then the dynamics of lambda lengths governed by the $A_{n-3}$ cluster algebra was merely the a dynamics for cluster variables. There exists however, a more general definition of cluster algebras where the coefficient variables are of dynamical type too\cite{Williams}. Then the reader can check in the literature (see Figure 8. of \cite{Williams} in this respect) that the coefficient dynamics found here is precisely the one of Zamolodchikov's $Y$-systems.
Notice that such systems originally have shown up in integrable deformations of conformal field theories in connection with thermodynamic Bethe ansatz equations for certain types of scattering problems\cite{Zamo}.
The exploration of this connection in the context of the boundary-bulk correspondence needs further elaboration.

Let us emphasize that the logarithms of $u_1$ and $u_2$ provide shears related to geodesic lengths of opposite sides of quadrangles as in Eq.(\ref{otherdist}). Moreover, these quantities under relations like (\ref{important}) are connected to conditional mutual informations. Hence the cluster algebraic structure found here is the bulk manifestation of gauge invariant entanglement patterns of the boundary.  

In summary lambda lengths provide coordinates for the decorated Teichm\"uller space of the bulk $\mathbb D$ with $n$ marked points on the boundary with a direct connection to regularized entanglement entropies. In this case the dynamics of regularized entropies is encapsulated in the cluster dynamics of an $A_{n-3}$ cluster algebra.
On the other hand their cross ratios provide coordinates for the Teichm\"uller space itself with a direct connection to conditional mutual informations. 
In this case the dynamics of this gauge invariant entanglement measure is encapsulated in the coefficient dynamics of the corresponding cluster algebra.

\section{An integral geometric analogy}

Finally we would like to link some of our obseravations to well-known results from integral geometry.
First of all notice that 
the plane waves of Eq.(\ref{helga}) giving rise to the (\ref{Wig}) expansions can be regarded as boundary to bulk propagators.
In order to see this notice that for $z=\tanh\frac{\varrho}{2}e^{i\varphi}$ and $\omega=e^{iu}$ one can write
\begin{equation}
\psi_k^{\pm}(z,u)=\left(\frac{1-\vert z\vert^2}{\vert z-\omega\vert^2}\right)^{\frac{1}{2}\pm ik}=\left[P(z,u)\right]^{\frac{1}{2}\pm ik}
\label{Poiker}
\end{equation}
where
$P(z,u)$ is the classical Poisson kernel.
Indeed the formula 
\begin{equation}\label{Poi}
\Psi(z)=\int_{\partial\mathbb D}P(z,\omega)c(\omega)d\omega,\qquad z\in{\mathbb D}
\end{equation}
with $d\omega=\frac{1}{2\pi}du$ gives the Poisson representation of a harmonic function\cite{Helga2}
in terms of the continuous boundary values encapsulated in the function $c(\omega)$.
In this notation the quantization of the geodesic motion manifests itself in the 
\begin{equation}\label{hujuj}
\Psi_k^{\pm}(z)=\int_{\partial\mathbb D}\left[P(z,\omega)\right]^{\frac{1}{2}\pm ik}c_k^{\pm}(\omega)d\omega,\qquad z\in{\mathbb D}
\end{equation}
future (sink) and past (source) boundary representations for the scattering states.

Let us consider the $g\in SU(1,1)$ transformation which takes the pair 
$(z_0,\omega_0)=(0,e^{iu_0})$ to
$(z,\omega)=(z,e^{iu})$.
Then a calculation\cite{Helga2} shows that
$\frac{du_0}{du}=P(z,\omega)$.
Hence for the (\ref{Poi}) Poisson transform we have
\begin{equation}
\int_{\partial{\mathbb D}}P(gz_0,\omega)c(\omega)d\omega =\int_{\partial{\mathbb D}}c(\omega)\frac{d(g^{-1}\omega)}{d\omega}d\omega
\end{equation}
Hence for $z=gz_0$ for the Poisson transform we have
\begin{equation}
\Psi(z)=(\mathcal{P}c)(gz_0)=\int_{\partial{\mathbb D}}c(g\omega)d\omega.
\end{equation}
In Ref.\cite{Helga3} it is shown that this boundary-bulk transformation gives rise to a Radon transform defined by a double fibration featuring the relevant dual coset spaces ${\mathbb D}=SU(1,1)/SO(2)$ and ${\partial \mathbb D}=SU(1,1)/H$ where $H$ is the stabilizer of the boundary point $(1,0)$.

For the scattering situation the density distributions of the sources and sinks $c_k^{\pm}(u)$ are related to the scattering state $\Psi(z)$ in a similar manner. The crucial difference that in this case we have\cite{BV}
\begin{equation}
c_k^{\pm\prime}(u^{\prime})=\left(\frac{du}{du^{\prime}}\right)^{\frac{1}{2}\mp ik}c_k^{\pm}(u)
\end{equation}
meaning that under fractional linear transformations these distributions transform as densities of order $\frac{1}{2}\mp ik$.

Let us now recall that for weakly exciting the vacuum in $AdS_3/CFT_2$ the change in the entanglement entropy of the region parametrized as $(u,v)=(\theta+\alpha,\theta-\alpha)$ is of the form\cite{Myers,Zukowski} 
\begin{equation*}
\delta S = 2\pi\int_{\theta-\alpha}^{\theta+\alpha}d\varphi\frac{\cos(\theta-\varphi)-\cos\alpha}{\sin\alpha}\langle T_{00}(\varphi)-T_{00}^{\rm vac}\rangle
\end{equation*}
where 
$\langle T_{00}^{\rm vac}\rangle=-\frac{c}{24\pi}$.
Using the coordinates of Eqs. (\ref{haha}) in this formula the integration kernel, regarded as a boundary-to-bulk propagator\cite{Myers} in $\mathbb K$, can alternatively be written as
\begin{equation}
Q(\gamma,\theta;\varphi)=\sinh\gamma+\cosh\gamma\cos(\theta-\varphi)
\end{equation}
which is the kinematic space analogue of the quantity showing up in Eq.(\ref{helga}). 
In this notation we can symbolically write
\begin{equation}\label{entropywave}
\delta S(w) = 2\pi\int_{I}Q(w,\varphi)\delta T(\varphi)d\varphi,\qquad w\in\mathbb K
\end{equation}
where for $w=(\gamma,\theta)$ we define the interval $I\subset\partial{\mathbb D}$ with boundary points defined by the pair $(\alpha+\theta,\alpha-\theta)$.
It is known\cite{Myers} that $\delta S$ obeys the Klein-Gordon equation with mass given by $m^2=-2$.

We can obtain an alternative understanding of this result by noticing that Eq.(\ref{entropywave}) can also be regarded as a special superposition of elementary solutions of a one dimensional Schr\"odinger equation with another form for a P\"oschl-Teller potential. Namely this Schr\"odinger equation is of the following form
\begin{equation}\label{sch2}
\left[-\frac{d^2}{d\gamma^2}-\frac{n^2-\frac{1}{4}}{\cosh^2\gamma}\right]u_n^j(\gamma)=-\left(j+\frac{1}{2}\right)^2u_n^j(\gamma).
\end{equation}
In order to see this take the Casimir operator $\triangle_{\mathbb K}=C_{\mathbb K}=-K_1^2-K_2^2+K_3^2$ with the operators ${\bf K}$ featuring Eq.(\ref{ittis}). Apply then a similarity transformation by $\sqrt{\cosh\gamma}$ to $C_{\mathbb K}$ then the Hamiltonian on the left hand side of Eq.(\ref{sch2}) can be written as $H=-(C_{\mathbb K}+\frac{1}{4})$. After acting on the ansatz $\psi(\gamma,\theta)=u_n^j(\gamma)e^{in\theta}$ 
and writing the eigenvalue of $C_{\mathbb K}$ in the form $j(j+1)$
one obtains (\ref{sch2}).
Now group theory tells us\cite{LN} that for a fixed $n$ integer or half integer the bound state spectrum of our Hamiltonian is given by
$E_j=-(j+\frac{1}{2})^2$ with $j=-1,-2,\dots -n$ or $j=-\frac{1}{2},-\frac{3}{2},\dots -n$. There is another discrete series of unitary irreducible representations but these describe the physically same situation due to the fact that our potential is symmetric under the exchange $n\mapsto -n$.

Now we recall at this point that originally we quantized the fast ($\mathbb H$) variables and we merely regarded the slow ($\mathbb K$) ones as parameters. Now it is clear that by writing up a Schr\"odinger equation of the (\ref{sch2}) form for the slow variables as well we
elevated them to the status of dynamical variables, amenable to quantization.
Since the Crofton form defines a natural symplectic and K\"ahler form on $\mathbb K$ the idea of also quantizing  the kinematic space degrees of freedom is a natural one\cite{Zuk2}.
Clearly in this picture in the spirit of Section II. we can regard Eq.(\ref{sch2}) this time as an equation arising from the quantization of the geodesic motion on kinematic space. 

The more familiar interpretation of our result in the usual holographic language is as follows.
One can check that $-\triangle_{\mathbb K}$ is the Klein-Gordon operator and the mass squared term is: $m^2=-j(j+1)$.
Hence the result of Ref.\cite{Myers} for $j=-2$ follows. Namely the $s=-j=2$ case corresponds to the situation of $T_{00}$ encapsulating boundary data with conformal weight $2$. Dually in the corresponding kinematic space the Lorentzian wave equation has tachionic mass, i.e. $m^2=-2$.
However, this holographic de Sitter construction also generalizes for conserved symmetric traceless currents $T_{\mu_1\mu_2\dots,\mu_s}$ with arbitrary spin\cite{Myers}.
In our notation we have $s=-j$ hence these currents in $d=2$ define charges $Q^{(s)}$ of the form
\begin{equation}\label{hajaj}
Q^{(s)}={(2\pi)}^{-j-1}\int_I\left[Q(w,\varphi)\right]^{-j-1} T_{00\dots 0}(\varphi)d\varphi.
\end{equation}
For $s=2$ we have $H_I=Q^{(2)}$ (modular Hamiltonian) 
then after taking expectation values via the first law of entanglement\cite{first}
$\delta S =\langle H_I\rangle$, and  
we get back to Eq.(\ref{entropywave}).
It is clear now that taking the expectation values on both sides of (\ref{hajaj}) results in an expression similar to Eq.(\ref{hujuj}).

In order to further compare the expressions (\ref{hujuj}) and (\ref{hajaj}) notice that the (\ref{sch2}) Schr\"odinger equation is dual to a one which should be familiar from Section V.
Indeed, this equation is the one arising from the dual Casimir operator $\triangle_{\mathbb H}=C_{\mathbb H}=-J_1^2-J_2^2+J_3^2$. In this case after similar manipulations one arrives at
\begin{equation}\label{schr3}
\left[-\frac{d^2}{d\varrho^2}+\frac{n^2-\frac{1}{4}}{\sinh^2\varrho}\right]v_n^j(\varrho)=-\left(j+\frac{1}{2}\right)^2v_n^j(\varrho).
\end{equation}
Formally one can relate these equations by analytic continuation $\gamma\mapsto \gamma-i\frac{\pi}{2}$ in (\ref{sch2}) and then replacing $\gamma$ with $\varrho$. 
However, in the second case ($\triangle_{\mathbb H}$) we have only the continuous representations $j=-\frac{1}{2}+ik$, $k\in{\mathbb R}^+$ since the (\ref{PT}) potential is repulsive and does not support bound states. These states give rise to the scattering results of Section V. 

On the other hand in the first case ($\triangle_{\mathbb K}$) we have scattering and bound states as well.
Now we have already revealed that the bound states with $j=-1,-2,\dots $ give rise to conserved charges of spin $s=-j$ which can be reinterpreted as scalar fields in $\mathbb K$ obeying a Lorentzian wave-equation.
Is there any obvious holographic interpretation of the scattering states of Eq.(\ref{sch2})?
In this respect we observe that (\ref{hujuj}) has two versions corresponding to the source and sink interpretations of the boundary distributions, on the other hand (\ref{hajaj}) seems to have merely one.
However, for pure states in the case of (\ref{hajaj}) we also have an antipodal symmetry which results in two classes of boundary data. This implies that just like in the $\mathbb H$ case we should have an intertwining relation in the scattering domain of the $\mathbb K$ case as well.
Recall that for the bound state case this antipodal map yields constraints (see Eq.(12) of Ref.\cite{Myers}) for the two types of boundary data. For the scattering domain this analogous relation should be similar to Eq.(\ref{relate2}) we used for calculating the scattering kernel. In order to clarify this point further insight is needed.

\section{Conclusions and comments}

In this paper we initiated a study of the space of horosurfaces in a holographic context.
Just like kinematic space $\mathbb K$, this space parametrizes a special family of submanifolds of $AdS_3$.
For the special case of a spacelike slice $\mathbb D$ these submanifolds are horocycles satisfying Eq.(\ref{horo}).
In this case there is a one to one correspondence between the space of horocycles, which we denoted by $\mathbb G$, and the points of ${\mathcal L}^{+}$, the positive light cone of $2+1$ dimensional Minkowski space.

Horocycles can be used to regularize geodesic lengths of  $\mathbb D$. Since regularized (lambda) lengths of bulk geodesics anchored to boundary intervals are related to their regularized entanglement entropies the space of horocycles is naturally related to the space of cut-offs. Regarding the freedom in choosing the cut-off as a gauge degree of freedom, one can say that the idea of horocycles geometrizes nicely the boundary gauge freedom in the bulk. The advantage of this dual approach is that the notion "the space of cut-offs" can be replaced by a mathematically well-defined object $\mathbb G$ which is a homogeneous space just like $\mathbb K$.

The physical basis which connects the spaces $\mathbb K$ and $\mathbb G$ is scattering theory.
According to Eq.(\ref{wavedistance}) suitably parametrized complex powers of regularized lambda lengths can naturally be regarded as scattering states of a Hamiltonian arising from the quantization of the classical geodesic motion on $\mathbb D$.
These scattering states are depending on both the coordinates of the bulk ($\mathbb D$) and kinematic space ($\mathbb K$). These coordinates can be regarded as fast and slow variables analogous to the standard Born-Oppenheimer treatment of electrons and nuclei. However, in our peculiar case this "fast-slow" parametrization is also linked to an arbitrary coice of a base point $z_0\in{\mathbb D}$ which is in turn connected to a special choice of the green horocycle of Figure 3 which is an element of $\mathbb G$.
In Eq.(\ref{Huygens1}) we have shown that this ambiguity in the choice of this horocycle is directly linked to a gauge degree of freedom. A convenient way to display these findings is the language of Berry's Phase.

As is well-known parametrized families of quantum states give rise to gauge structures living on parameter space.
In our case the parameter space corresponds to the space of slow variables which is kinematic space $\mathbb K$, and the families of quantum states are our scattering states. We have shown that the Berry curvature of the connection responsible for the $GL(1,\mathbb C)$ holonomy of the previous paragraph, is just the Crofton form with a coefficient function depending on the scattering energy.
This is what was expected since $\mathbb K$ is a symmetric space of the form $G/H$ and in this case Berry's connection is simply related to the Riemannian connection on $\mathbb K$, provided that the matrix elements of the generators not belonging to $H$ are vanishing. The latter condition holds for our case of scattering states.
Another consequence of this is the fact that the symmetric part of the quantum geometric tensor\cite{Provost} is up to a scattering energy dependent function is just the metric of Eq.(\ref{PVmetr}) known form \cite{Czech1}.   
It would be of some interest of calculating this function and identify its meaning in a holographic context. 

Notice also that the fact that Berry's curvature is proportional to the Crofton form, should really be spelled out differently. Indeed, this curvature form should rather be regarded as a one proportional to the natural Kirillov-Kostant symplectic form defined on a coadjoint orbit. This idea is well-known from the literature on geometric quantization. Since our geodesic operator of Eq.(\ref{coadjoint}) underlying our calculations is of coadjoint type, our result is in accord with other ones written in this spirit\cite{Vinet}. 
As far as a link with these observations and holographic duality is concerned notice that kinematic space is just a particular coadjoint orbit of the conformal group $SO(d,2)$, with the Crofton form for $d=2$ equals the standard Kirillov-Kostant form\cite{Zukowski}.
Since $\mathbb K$ is a symplectic manifold, it can be quantized according to the methods of geometric quantization.
In our analogy with fast and slow variables, that would mean to regard our space of parameters $\mathbb K$ as dynamical variables also amenable to quantization.
Some useful piece of information has already shown up in this respect in \cite{Zukowski} where observations on the possibility of also quantizing the kinematic space variables were presented. 
Since kinematic space is the single sheeted hyperboloid, quantization via the orbit method produces naturally the principal series representation of the conformal group. This is precisely the representation, labelled by the wave number $k$, taking care of our scattering states coming from the quantization of geodesic motion. It would be interesting to clarify within this framework the physical meaning of the $k\in{\mathbb R}^+$ dependent prefactor of the Crofton form showing up in Eq.(\ref{curvature}).    

In Section V. we observed that our scattering states based on lambda length constructions provide two different types of expansions for an arbitrary scattering state. They are associated with two different boundary distributions corresponding to the starting and end points of the geodesics. These expansions encapsulate superpositions of plane waves absorbed or emitted at all boundary points. These future and past representations are related by a scattering operator. We observed that the derivative of the phase of the kernel of this operator with respect to the scattering energy is proportional to the entanglement entropy.
This result relates a Wigner delay type quantity to the entanglement entropy.
Though formally this construction can be generalized to more general scattering scenarios by the method of intertwiners\cite{Kerimov}, it is not obvious to us whether a physically sound relationship between Wigner delays and entanglement entropies found here can also be established for more general holographic situations. 

In Sections VI.-VII. by making use of the gauge degree of freedom provided by horocycles we reconsidered standard results on strong subadditivity using lambda lengths.
The interesting new possibility showing up in our treatise was the use of formula (\ref{RTlambda2}) in holographic considerations, making full use of the space $\mathbb G$.
This approach culminating in Eqs.(\ref{ssshear4}) have revealed a connection between conditional mutual information
and shear coordinates parametrizing the deformation (Teichm\"uller) space of geodesic quadrangles. For infinitesimal quadrangles one gets back to the result of \cite{Czech1} in a new (\ref{ssshear3}) form. 
As a byproduct of our approach (see Eq.(\ref{important})) a dual interpretation of boundary conditional mutual information in terms of bulk geodesic distances of opposite sides of hyperbolic quadrangles emerged.
We also explored the kinematic space representation of these results summarized in expressions (\ref{kincausal})-(\ref{shearRT}).
Generalizing these ideas we have shown that that for geodesic polygons the chain rule for conditional mutual information expressing the fact that volumes in kinematic space are additive, can be regarded as ones arising from the (66) patching conditions for shear coordinates.

We have emphasized that the basic role the gauge invariant conditional mutual informations play in these elaborations 
dates back to the unifying role of the Ptolemy identity (\ref{ptolemy}).
This identity is the basis of recursion relations underlying transformation formulas like the ones we have obtained for the (\ref{u1u2}) pair $(u_1,u_2)$ of shear coordinates of geodesic pentagons. 
Indeed in the general case of geodesic $n$-gons these recursion relations are precisely of the form of Zamolodchikov's $Y$-systems of $A_{n-3}$ type. A consequence of this is that for the vacuum state of the CFT the $A_{n-3}$ cluster algebra provides the algebraic structure for organizing 
the gauge invariant entanglement patterns of the boundary into the bulk.
Note that for this to work the more general formulation of cluster algebras is needed where the coefficient variables are also of dynamical type\cite{Williams}. We conjecture that a holographic interpretation of the corresponding coefficient dynamics answering the dynamics of Zamolodchikov $Y$-systems is directly related to the dynamics of holographic codes
discussed in Ref.\cite{Osborne}. Indeed, for the construction of such dynamics Pachner moves\cite{Osborne} and the Ptolemy grupoid play a basic role. Such moves are just the flips underlying our cluster dynamics and the mutation of the corresponding quivers\cite{Williams} is just its representation in kinematic space.
One can convince oneself on this point by having a look at Figure 3. of Ref.\cite{Williams} showing that triangulations of geodesic $n$-gons in $\mathbb D$ give rise to quivers in $\mathbb K$. Explicitely: putting a frozen vertex at the midpoint of the $n$ sides and a mutable vertex at the midpoint of the $n-3$ diagonals of an $n$-gon labelled by the corresponding vectors of $\mathbb K$ gives rise to a quiver reminiscent of a tensor network living in kinematic space. 
An elaboration of these ideas would connect the algebraic structure of cluster algebras to current research topics like holographic codes\cite{Yoshida},  MERA\cite{Czechmera} etc. 

In Section VIII. we embarked in an elaboration of the integral geometric implications of our results.
Eq.(\ref{Poiker}) of this section is clearly illustrating that quantities like the one of (\ref{wavedistance}), based on lambda lengths connected to our space $\mathbb G$, are related to integral kernels. In particular the future and past representations for scattering states of Eq.(\ref{hujuj}) are just Radon-like transforms between the bulk and the boundary based on the corresponding double fibration.
The elementary scattering problem associated to this situation is given in terms of 
$\triangle_{\mathbb D}$
in Eq.(\ref{schr3}).
Interestingly there is an analogous scattering problem of Eq.(\ref{sch2}) associated to the situation of weakly exciting the vacuum in $ADS_3/CFT_2$ featuring $\triangle_{\mathbb K}$.
Such scattering problems on $\mathbb D$ and $\mathbb K$ are related by analytic continuations.
Let us remark here that interestingly both of the scattering potentials of Eqs.(\ref{schr3}) and (\ref{sch2}) show up in the scattering problem of Ref.\cite{Perry} occurring in connection with the trace formula for the Euclidean form of the BTZ black hole metric.

In this paper we have not attempted to dwell in the integral geometric subtleties of the double fibration featuring our basic actor: the space $\mathbb G$. This would mean working out the physical implications of a horocycle transformation\cite{Helga3} between $\mathbb D$ and $\mathbb G$, which is analogous to the X-ray transform used succesfully in Refs.\cite{Czech1,Czech1b}. The exploration of such ideas we postpone for future work.    

In our work we explored the simplest instance of gauge structures related to the entanglement patterns of the CFT vacuum.
We achieved this task by studying the scattering states, parametrized by the coordinates of $\mathbb K$, associated to quantized geodesic motion.
The obvious question is, how to generalize our findings to more general CFT states reflecting other holographic scenarios?
The simplest option for exploring the gauge structure of other boundary states is to study their dual classical geometries in the bulk with a different type of geodesic motion to be quantized.
Since in $2+1$ dimensional gravity there are no local gravitational degreees of freedom in the bulk side of the $AdS_3/CFT_2$ correspondence global issues are of importance. This means that since every classical solution in $AdS_3$ gravity is locally $AdS_3$ then all the interesting global geometries to be studied in this context are
of the form $AdS_3/\Gamma$
for some discrete subgroup $\Gamma$ of $SO(2,2)$.
Such solutions can be thought of as generalized eternal BTZ black holes\cite{BTZ,Ingemar1,Brill,Skenderis}.
In this picture for a static spacetime the simplest example when $\Gamma$ is a Fuchsian group of $PSL(2,\mathbb R)$ generated by a fundamental element the corresponding geometry is the BTZ black hole which is a two boundary wormhole.
Choosing more interesting examples for $\Gamma$ yields multi-boundary wormholes which has already been investigated in the holographic context\cite{Skenderis}.
Such spacetimes can be obtained by factorising $\mathbb U$ by a subgroup $\Gamma$ of  $PSL(2,\mathbb R)$ obtaining a Riemann surface $S$. The upper half plane $\mathbb U$ is then embedded into $AdS_3$ and the action of $\Gamma$ is extended to $AdS_3$\cite{Brill,Skenderis}.
For the static slice then one should consider the scattering states of the corresponding quantized geodesic motion and check how the entanglement patterns of the boundary manifest themselves in scattering data, like Wigner delays, scattering matrices etc. 

For {\it extremal} multiboundary\cite{Ingemar1,Brill} wormholes this means studying scattering problems similar to the ones familiar from the literature on quantum chaos\cite{Gutz,BV,Comtet,Pnueli}. 
These investigation were based on classical results on the scattering theory of automorphic functions\cite{Fad,LP} based on Eisenstein series.
One particular example of that kind is Gutzwiller's leaky torus\cite{Gutz}, which is a Riemann surface with genus one, with a cusp. 
The complication with the idea of using such examples in a holographic context is that what one really has to consider is not merely a scattering problem but a parametrized family of such problems where the parameter space is either the Teichm\"uller space of $S$, or a fundamental domain of kinematic space\cite{Zhang}.
The other problem is that for models with physical import one should leave the domain of extremal wormholes. For non-extremal multiboundary wormholes no results of that kind to be used are known to the author.
In this context we must note however, that these complications can be evaded for
the scattering situation showing up for the Euclidean form of the BTZ black hole metric in Ref.\cite{Perry}. As we have already emphasized, here the
 scattering situation is simply coincides with a one featuring both of the potentials of Eqs.(\ref{schr3}) and (\ref{sch2}).
There is a caveat however, since the potential strength parameter $n^2$ of Eq.(\ref{sch2}) in the BTZ context should be changed to $-n^2$. In any case revisiting the approach adopted in that paper could be the first step for generalizing our scattering related considerations for a situation with $AdS_3/\Gamma$ type geometry encoding more sophisticated quantum states then the CFT vacuum.

In Section VII. we have shown that for the $CFT_2$ vacuum the $A_{n-3}$ cluster algebra plays an important role in describing how the gauge invariant conditional mutual informations of intersecting boundary intervals, giving rise to geodesic $n$-gons, can be patched together using shear coordinates of elementary bulk quadrangles. Moreover, we have observed that the cluster dynamics based on flips (or alternatively on quiver mutations) provides a dynamics similar to the ones conjectured for holographic codes\cite{Osborne}.
How these observation can be generalized for more general $CFT_2$ states? In order to answer this question the simplest example to be explored would be the BTZ black hole solution which is dual to a thermal state of the corresponding $CFT_2$.
It is known that the $t=0$ geometry of the extremal BTZ solution is that of a punctured disk (see Figure 7. of Ref.\cite{Brill}).              
Moreover, it is also known\cite{Fomin} that  for $n\geq 4$, $n$-gons with a puncture give rise to a cluster algebra of type $D_n$.
In particular the simplest $n=4$ case of quadrangles of the punctured disk is connected to the coordinate ring of $Gr(6,3)$ the Grassmannian of three-planes through the origin in a six dimensional vector space\cite{Williams,Fomin}.
This observations indicate that there could be a natural way to associate different types of cluster algebras to a particular subclass of CFT states. As another example giving some support to this expectation one can also notice that according to Figure 7 of Ref.\cite{Brill} the $t=0$ geometry of the nonextremal BTZ should be equivalent to an annulus. Then one can consider the situation of a pair $(n_1,n_2)$ of marked points lying on the corresponding boundaries. By considering triangulations the corresponding cluster algebra is the one listed in Table 1. of Ref.\cite{Fomin}.

In the most general case in the context of Ref.\cite{Skenderis} one should then start with an arbitrary bordered Riemann surface $S$ with a set $M$ of $n$ marked points. One can then consider ideal triangulations of $(S,M)$ and their associated lambda lengths a setup which provides cluster algebraic structures associated to $(S,M)$ in a natural manner. Actually this is the original context where cluster algebras have shown up\cite{Williams}.
One can even guess that for multiboundary wormholes, cluster algebras might provide a natural algebraic means for encoding the gauge invariant entanglement patterns of a certain class of boundary entangled states in the geometry of bulk geodesics.
Moreover, the corresponding cluster dynamics might  educate us on dynamical issues concerning holographic codes.
Since the corresponding quiver dynamics based on mutation can naturally be represented in the corresponding kinematic spaces
this could provide us with a natural algebraic setup for generalizing the observations on the MERA network living in $\mathbb K$ associated to the CFT vacuum\cite{Czechmera}. In order to explore this possibility further, one should start trying to set up a dictionary between the mathematical structures provided by tensor networks and holographic codes on one side and the ones of cluster algebras on the other.
This interesting idea needs further elaboration.

\section{Acknowledgement}

The author would like to express his gratitude to Zsolt Szab\'o for his work with the figures of this paper.
This work was supported by the National Research Development and Innovation Office of Hungary within the Quantum Technology National Excellence Program (Project No. 2017-1.2.1-NKP-2017-0001).

\section{AppendixA: Horocycles and lambda lengths}

We consider special curves in ${\mathbb H}$ called horocycles.
Let us denote by ${\mathbb R}^{2,1}$ the $2+1$ dimensional Minkowski space-time equipped with the Minkowski inner product. Define the positive light-cone $\mathcal{L}^+\subset {\mathbb R}^{2,1}$ as the set of vectors ${\bf b}$ satisfying
${\bf b}\cdot{\bf b}\equiv b_1^2+b_2^2-b_3^2=0$ and $b_3>0$.
Then a {\it horocycle} $h$ is a curve with points $X\in{\mathbb H}\subset {\mathbb R}^{2,1}$ (see Eq.(1) for notation) satisfying
\begin{equation}
{\bf X}\cdot{\bf b} =-1/\sqrt{2},\qquad {\bf X}\cdot {\bf X}=-1,\qquad {\bf b}\cdot{\bf b} =0
\label{horo}
\end{equation}
where for the rationale of the appearance of the number $1/\sqrt{2}$ see Ref.\cite{Pennerbook}.
Note that the correspondence between the set of horocycles and the set of points in $\mathcal{L}^+$ is one to one. 
In the text we denote the space of horocycles by either of the symbols $\mathbb G$ and $\mathcal{L}^+$. 
The first of them ($\mathbb G$) refers to its meaning as the space of gauge choices corresponding to the the gauge-like degrees of freedom showing up in the Berry's Phase context of Section IV. 

From Eq.(\ref{horo}) it follows that when viewed in ${\mathbb D}$ the points $(x,y)$ lying on the horocycle satisfy the equation
\begin{equation}
\left(x-\frac{\sqrt{2}b_1}{1+\sqrt{2}b_3}\right)^2+\left(y-\frac{\sqrt{2}b_2}{1+\sqrt{2}b_3}\right)^2 =\frac{1}{(1+\sqrt{2}b_3)^2}
\end{equation}
where $x=\frac{X}{1+U}$ and $y=\frac{Y}{1+U}$. 
The light-like vector ${\bf b}\in{\mathbb R}^{2,1}$ also defines a point $\omega$ in the boundary with coordinates $(\omega_1,\omega_2)\equiv (b_1/b_3,b_2/b_3)\in{\partial}{\mathbb D}$.
Clearly the image of the horocycle in $\mathbb D$ is a Euclidean circle with radius $r=1/{(1+\sqrt{2}b_3)}$
tangent to the boundary at $\omega$ with $2r$ as the penetration depth of the horocycle. 
One can make contact with the notation used in the considerations following Eq.(\ref{generalplane}) by noting that
\begin{equation}
\omega=e^{iu}\equiv \omega_1+i \omega_2=\frac{b_1+ib_2}{b_3}\in\partial{\mathbb D}.
\end{equation}

Let us now consider a geodesic departing from $\omega_- =e^{iu}$ and arriving at $\omega_+ =e^{iv}$.
Let the corresponding light-like vectors of ${\mathbb R}^{2,1}$ denoted by ${\bf b}^{\pm}$.
They define horocycles $h_+$ and $h_-$.
We have ${\bf b}^+\cdot{\bf b}^- =b_1^+ b_1^-+ b_2^+ b_2^--b_3^+ b_3^-$.

Now we define the {\it lambda length} of the geodesic labelled by the pair $(u,v)\in{\mathbb K}$ as
\begin{equation}
\lambda(h_+,h_-)\equiv \sqrt{-{\bf b}^+\cdot{\bf b}^-}.
\label{lambdalength}
\end{equation}
Then we have the result\cite{Penner,Pennerbook}
stating that the signed Poincar\'e distance $d(h_+,h_-)$ along the geodesic from $\omega_-$ to $\omega_+$ between $h_-$ and $h_+$, taken with positive sign if $h_+\cap h_-$ is disjoint and with negative sign otherwise, is related to the lambda length as
\begin{equation}
\lambda^2(h_+,h_-)=e^{d(h_+,h_-)}.
\label{lambdalength2}
\end{equation}

Fixing a point of departure $\omega\equiv\omega_-$ for an oriented geodesic in the boundary and another point $z_0$ in the bulk lying on it fixes a horocycle $h_-(z_0)$ and the point of arrival $\omega_+$. Moreover, for a varying point $z$ in the bulk the corresponding horocycles $h_+(z)$ are also fixed uniquely.
Then we have
\begin{equation}
\lambda^2(h_+(z),h_-(z_0))=e^{d(h_+(z),h_-(z_0))}=e^{d(\omega;z_0,z)}
\label{kapcs}
\end{equation}
clarifying the meaning of Eq.(\ref{wavedistance}) in terms of the lambda length.

As an example take
\begin{equation*}
{\bf b}^{\pm}\in{\mathcal L}^+\leftrightarrow \frac{1}{\sqrt{2}}(\pm 1,0,1)
\end{equation*}
where the penetration depths are $2r_+=2r_-=1$. These data give rise to the diametrical geodesic of Figure 2. with end points points $(\pm 1,0)\in\partial\mathbb D$. Then (unlike in Figure 2.) the corresponding horocycles are touching each other at the origin of $\mathbb D$ yielding for the Poincar\'e length of the geodesic segment $d(h_+,h_-)=0$ in accordance with $-{\bf b}^{+}\cdot {\bf b}^{-}=1=e^{d(h_+,h_-)}$. Hence for this special choice the lambda length is unity.
Notice that as an alternative description for $\mathbb G$ can be given as the set of $SO_0(2,1)$ orbits of the distinguished horocycle defined by ${\bf b}^{+}$.

Note that in the upper half plane model the boundary $\partial{\mathbb U}$ is 
${\mathbb R}{\mathbb P}^1 ={\mathbb R}\cup\{i \infty\}$.
For a segment $A$, with Euclidean length $L_A$, belonging to the ${\mathbb R}$ part of the boundary the lambda length of $C$ is given by the simple formula\cite{Pennerbook}
\begin{equation}
\lambda(A)=\frac{L_A}{\sqrt{\Delta_b\Delta_c}}=e^{d(\tau_b,\tau_c)/2},\qquad L_A=\vert\xi_c-\xi_b\vert
\label{simplelambda}
\end{equation}
where $\Delta_b$ and $\Delta_c$ are the Euclidean lengths for the diameters of the corresponding horocycles, and $\tau_b,\tau_c\in {\mathbb U}$ are the endpoint coordinates of the geodesic segment lying in between the corresponding horocycles.
This formula is valid for any geodesic which is represented as a circular arc centered on ${\mathbb R}$. The formula is illustrated in Figure 6.
\begin{figure}

\centerline{\includegraphics{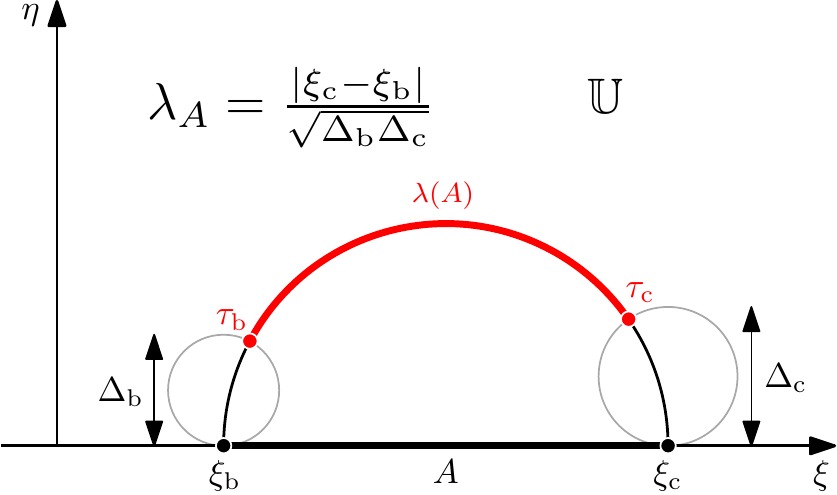}}
\caption{Illustration of the meaning of the lambda length for a circular arc centered on ${\mathbb R}$ which is a part of the boundary $\partial{\mathbb U}$.}
\end{figure}
For the other type of geodesics which are just straight lines parallel to the $\eta$ axis for the lambda length
we have\cite{Pennerbook}
\begin{equation}
\lambda =\sqrt{\frac{\eta}{\Delta}}
\label{simplelambda2}
\end{equation}
i.e. it is showing merely dependence on the parameters $\eta$ and $\Delta$ characterizing the horocycles centered at ${i\infty}$ and an arbitrary point on ${\mathbb R}$.  

Now we relax the assumption of taking merely the static slice, and define lambda lengths for $AdS_3$.
In order to do this we elevate the three component vectors of Eq.(\ref{horo}) taken from ${\mathbb R}^{2,1}$ to the status of four component ones taken from ${\mathbb R}^{2,2}$. Hence now $\bf X$ and ${\bf X}^{\prime}$  has components $(X,Y,U,V)$ and $(X^{\prime},Y^{\prime},U^{\prime},V^{\prime})$ and 
\begin{equation}
{\bf X}^{\prime}\cdot{\bf X}=X^{\prime}X+Y^{\prime}Y-U^{\prime}U-V^{\prime}V.
\label{minkskal}
\end{equation}
In this notation $AdS_3$ is the embedded surface given by the one sheeted hyperboloid ${\bf X}\cdot {\bf X}=-1$.
Alternatively one can define the matrix
\begin{equation}
\hat{\bf X}\equiv\begin{pmatrix}U+X&Y-V\\Y+V&U-X\end{pmatrix}
\end{equation}
then $AdS_3$ is defined by the equation ${\rm Det}\hat{\bf X}=1$. In this picture $AdS_3$ is the group manifold $SL(2,\mathbb R)$
and the connected part of the isometry group $SO(2,2)\simeq SL(2,\mathbb R)\times SL(2,\mathbb R)/{\mathbb Z}_2$
acts as $\hat{\bf X}\mapsto g_L\hat{\bf X}{g_R}^T$ with  $g_L,g_R\in SL(2,{\mathbb R})$.
Moreover, for two points of ${\mathbb R}^{2,2}$ we have
\begin{equation*}
{\bf X}^{\prime}\cdot{\bf X}=\frac{1}{2}{\rm Tr}(\hat{\bf X}^{\prime}\varepsilon\hat{\bf X}^T\varepsilon)
\end{equation*}
where $\varepsilon$ is the $2\times 2$ antisymmetric tensor with $\varepsilon_{12}=-\varepsilon_{21}=1$.

For lambda length calculations we are interested in spacelike separated points $\hat{\bf X}^{\prime},\hat{\bf X}\in SL(2,{\mathbb R})$. In the special case when $\hat{\bf X}^{\prime}=I$ this implies that ${\rm Tr}(\hat{\bf X})>2$, i.e. $U>1$.
In analogy to Eqs.(\ref{horo}) one can regard the vectors 
${\bf b}^{\pm}$ as four component null vectors, or alternatively rank one matrices $\hat{\bf b}^{\pm}$.
Then one can write
\begin{subequations}\label{minkrank}
\begin{align}
\hat{\bf b}^{+}={\bf u}_L{{\bf u}_R}^T\\
\hat{\bf b}^{-}={\bf v}_L {{\bf v}_R}^T
\end{align}
\end{subequations}
where the notation refers to the dyadic product of two two-component column vectors.

Now the four component analogues of Eqs.(\ref{horo}) define objects we call horosurfaces
\begin{equation*}
h_{\pm}=\{\hat{\bf X}\in SL(2,\mathbb R)\vert\quad {\bf X}\cdot {\bf b}^{\pm}=-1/\sqrt{2},\quad {\bf b}^{\pm}\cdot
{\bf b}^{\pm}=0\}
\end{equation*}
with ${\rm Tr}(\hat{\bf X})>2$, ${\rm Tr}(\hat{\bf b}^{\pm})>2$.
As in the proof of Lemma 2.1 of Ref.\cite{Penner} we can homogenize the defining equation of horosurfaces as
$2({\bf X}\cdot {\bf b}^{\pm})^2=-{\bf X}\cdot{\bf X}$. Then for space-like geodesics we have
\begin{equation}
\cosh^2 d({\bf X},{\bf X}^{\prime})=\frac{({\bf X}^{\prime}\cdot{\bf X})^2}{{\bf X}^2{{\bf X}^{\prime}}^2}.
\label{homodist}
\end{equation}
Demanding that
\begin{subequations}
\begin{align}
{\bf X}=t{\bf b}^++(1-t){\bf b}^-\in h_+\\
{\bf X}^{\prime}=s{\bf b}^++(1-s){\bf b}^-\in h_-
\end{align}
\end{subequations} 
we obtain $s=1-t=(1+\lambda^2)^{-1}$ where $\lambda^2(h_+,h_-)=-{\bf b}^+\cdot{\bf b}^-$.
Using this in Eq.(\ref{homodist}) we obtain the result
\begin{equation}
e^{d(h_+,h_-)}=\lambda^2(h_+,h_-)=-{\bf b}^+\cdot{\bf b}^-
\label{hiperfontos}
\end{equation}
which is just the straightforward generalization of Eq.(\ref{lambdalength2}).

Using Eqs.(\ref{minkskal})-(\ref{minkrank}) one can also obtain an the alternative formula
\begin{equation*}
e^{d(h_+,h_-)}=\lambda^2(h_+,h_-)=\frac{1}{2}({{\bf u}_R}^T\varepsilon {\bf v}_R)({{\bf u}_L}^T\varepsilon {\bf v}_L).
\label{hiperfontos2}
\end{equation*}
Finally for the geodesic length between two horosurfaces one obtains
\begin{equation*}
d(h_+,h_-)=\log({{\bf u}_R}^T\varepsilon {\bf v}_R)({{\bf u}_L}^T\varepsilon {\bf v}_L)-\log 2=\ell_{reg}-\log2.
\end{equation*}
where $\ell_{reg}$ is the regularized length between boundary points introduced in Ref.\cite{Maxfield}.
Hence this argument relates  $\ell_{reg}$  to a generalization of Penner's lambda length.
For the static $V=0$ slice $\hat{\b X}$ is a symmetric matrix hence ${\bf u}\equiv{\bf u}_L={\bf u}_R$ and ${\bf v}\equiv {\bf v}_L={\bf v}_R$ in this case the formula gives an alternative expression of the usual lambda length.
Moreover, for the quadrangles of Section VI. the four boundary points can be described by the four two-component vectors ${\bf u},{\bf v},{\bf w}$ and ${\bf z}$. The eight components of these column vectors can be arranged in a $2\times 4$ matrix $({\bf u}\vert{\bf v}\vert{\bf w}\vert{\bf z})$ with its six possible minors $P_{\mu\nu}$ with $\mu,\nu=1,2,3,4$ serve as Pl\"ucker coordinates for the Grassmannian $Gr(2,4)$.  
For example one has $P_{12}={\bf u}^T\varepsilon {\bf v}$. In this picture the Pl\"ucker relation $P_{12}P_{34}+P_{14}P_{23}=P_{13}P_{24}$ boils down to the Ptolemy relation of Eq.(\ref{ptolemy}) the basic relation of the $A_1$ cluster algebra\cite{Williams}.

\section{Appendix B: Double Fibrations}

Here for the convenience of the reader we summarize the set of double fibrations based on the group $G=SU(1,1)$ implicitely used in the text.
We follow the notation of Chapter I. of \cite{Helga3}.

Let us consider the subgroups of $G$
\begin{equation*}
K=SO(2)=\left \{\begin{pmatrix}e^{i\varphi}&0\\0&e^{-i\varphi}\end{pmatrix}: \varphi\in\mathbb {R}\right \}
\end{equation*}
\begin{equation*}
A=SO(1,1)=\left \{\begin{pmatrix}\cosh\gamma&\sinh\gamma\\\sinh\gamma&\cosh\gamma\end{pmatrix}: \gamma\in\mathbb {R}\right \}
\end{equation*}
\begin{equation*}
M={\mathbb Z}_2=\left \{\begin{pmatrix}\varepsilon&0\\0&\varepsilon\end{pmatrix}: \varepsilon^2=1\right \}
\end{equation*}
\begin{equation*}
M^{\prime}=\left \{\begin{pmatrix}e^{i\varphi}&0\\0&e^{-i\varphi}\end{pmatrix}: \varphi=0,\pm\frac{\pi}{2},\pi\right \}
\end{equation*}
\begin{equation*}
N=\left \{\begin{pmatrix}1+it&-it\\it &1-it\end{pmatrix}: t\in\mathbb {R}\right \}.
\end{equation*}
Let us choose $H= M^{\prime}A$ and $L=K\cap H$.
Then in this notation we have the two maps $\pi_1:G/L\to \mathbb D$ and $\pi_2: G/L\to \mathbb K^{\prime}$ where
\begin{equation*}
\mathbb D =G/K,\qquad {\mathbb K}^{\prime}=G/H
\end{equation*}
i.e. the Poincar\'e disc regarded as the spacelike slice of $AdS_3$ and its space of unoriented geodesics ${\mathbb K}^{\prime}$
forms a double fibration of the group $G$. 
This latter term means that $\pi_{1,2}$ are projections of the respective fibre boundles, the map $\pi_1\times \pi_2:G\to \mathbb D\times \mathbb K^{\prime}$ is an immersion, and moreover for each $z\in\mathbb D$ and $\zeta\in\mathbb K^{\prime}$ the sets ${\mathbb D}_{\zeta}\equiv \pi_1(\pi_2^{-1}(\zeta))\subset \mathbb D$ (geodesics of $\mathbb D$ parametrized by the points of $\mathbb K^{\prime}$) and
${\mathbb K^{\prime}}_{z}\equiv \pi_2(\pi_1^{-1}(z))\subset \mathbb K^{\prime}$
(the point curves of $\mathbb K^{\prime}$ parametrized by the points of $\mathbb D$) are smooth submanifolds.
It can be shown that $G/L$ can naturally be identified with the space of pairs of left cosets of the form
$(gK,gH)$ where $gK\cap gH\neq 0$. Clearly fixing either coset the latter relation can be regarded as an incidence relation defining point curves and geodesics of the respective spaces.
As explained in Refs.\cite{Czech1,Czech1b} the geometric data on the spaces $\mathbb D$ and $\mathbb K^{\prime}$ is connected by the $X$-ray transform.

Let us now choose $H^{\prime}=MN$ and $L^{\prime}=K\cap H^{\prime}$.
Then in this case in an analogous manner we have a corresponding double fibration between the two spaces
\begin{equation*}
\mathbb D =G/K,\qquad \mathbb G=G/H^{\prime}
\end{equation*}
which gives a correspondence between the static slice of the bulk and the space of horocycles. The relevant transformation in this case is the {\it horocycle transform}.

Finally one can choose $H^{\prime\prime}=MAN$ and $L^{\prime\prime}=K\cap H^{\prime\prime}$.
In this case $MAN$ is the subgroup leaving invariant the point $(1,0)$ of $\mathbb D$.
The corresponding double fibration is between the two spaces
\begin{equation*}
\mathbb D =G/K,\qquad \partial{\mathbb D}=G/H^{\prime\prime}
\end{equation*}
which gives a bulk-boundary correspondence. The integral transform associated with this situation is the classical Poisson transform. A version of this transformation have been used in Eq.(\ref{hujuj}) of Section VIII. where we transformed the boundary distributions of sources and sinks to the two different representatives of the scattering wave functions.
As discussed in Section V. the intertwining relation between these two representatives gives rise to the kernel of Eq.(\ref{kernelfv}) used in our considerations of the Wigner delay.

\end{document}